\documentclass[article,twocolumn,pra,shortbibliography,showpacs,superscriptaddress,nofootinbib]{revtex4-1}

\usepackage{graphicx}
\usepackage{bm}
\usepackage{amssymb}
\usepackage{ulem}
\usepackage{enumitem}
\usepackage{color}
\usepackage{amsmath}
\usepackage{amstext}
\usepackage{latexsym}
\usepackage[title]{appendix}
\usepackage[usenames,dvipsnames]{xcolor}
\usepackage[colorlinks=true,citecolor=blue,linkcolor=RubineRed,urlcolor=blue]{hyperref}
\usepackage{multirow}
\renewcommand{\thesection}{\Roman{section}} 

\renewcommand{\thesubsection}{\Alph{subsection}}
\usepackage{phfqit}
\usepackage{dsfont}
\usepackage{txfonts}

\usepackage{hyperref}%
\colorlet{linkcolor}{blue}
\hypersetup{bookmarksnumbered=false,bookmarksopen=false,bookmarksopenlevel=1,%
  breaklinks=true,pdfborder={0 0 0},colorlinks=true,%
  anchorcolor=linkcolor,citecolor=linkcolor,%
  filecolor=linkcolor,linkcolor=linkcolor,%
  menucolor=linkcolor,runcolor=linkcolor,%
  urlcolor=linkcolor}
\usepackage[all]{hypcap}%

\usepackage[nameinlink,capitalize]{cleveref}
\crefname{thm}{Theorem}{Theorems}

\begin{document}

\title{Fractional resonances and prethermal states in Floquet systems}

\author{R. Pe\~{n}a}
\affiliation{Departamento de F\'isica, Universidad de Santiago de Chile, 
Avenida V\'ictor Jara 3493, 9170124, Santiago, Chile}

\author{V. M. Bastidas}
\affiliation{NTT Basic Research Laboratories and Research Center for Theoretical Quantum Physics, 3-1 Morinosato-Wakamiya, Atsugi, Kanagawa 243-0198, Japan}
\affiliation{National Institute of Informatics, 2-1-2 Hitotsubashi, Chiyoda-ku Tokio 101-8430, Japan}

\author{F. Torres}
\affiliation{Departamento de F\'isica, Facultad de Ciencias,
Universidad de Chile, Casilla 653, Santiago, Chile 7800024}
\affiliation{Center for the Development of Nanoscience and Nanotechnology, Estaci\'on Central, 9170124, Santiago, Chile}
\affiliation{Department of Physics, University of California, San Diego, La Jolla, CA 92093, USA}

\author{W. J. Munro}
\affiliation{NTT Basic Research Laboratories and Research Center for Theoretical Quantum Physics, 3-1 Morinosato-Wakamiya, Atsugi, Kanagawa 243-0198, Japan}
\affiliation{National Institute of Informatics, 2-1-2 Hitotsubashi, Chiyoda-ku Tokio 101-8430, Japan}

\author{G. Romero}
\affiliation{Departamento de F\'isica, Universidad de Santiago de Chile, 
Avenida V\'ictor Jara 3493, 9170124, Santiago, Chile}
\affiliation{Center for the Development of Nanoscience and Nanotechnology, Estaci\'on Central, 9170124, Santiago, Chile}

\date{\today}

\begin{abstract}
In periodically-driven quantum systems, resonances can induce exotic nonequilibrium behavior and new phases of matter without static analog. We report on the emergence of fractional and integer resonances in a broad class of many-body Hamiltonians with a modulated hopping with a frequency that is either a fraction or an integer of the on-site interaction. We contend that there is a fundamental difference between these resonances when interactions bring the system to a Floquet prethermal state. Second-order processes dominate the dynamics in the fractional resonance case, leading to less entanglement and more localized quantum states than in the integer resonance case dominated by first-order processes. We demonstrate the dominating emergence of fractional resonances using the Magnus expansion of the effective Hamiltonian and quantify their effects on the many-body dynamics via quantum states' von Neumann entropy and Loschmidt echo. Our findings reveal novel features of the nonequilibrium quantum many-body system, such as the coexistence of Floquet prethermalization and localization, that may allow to development of quantum memories for quantum technologies and quantum information processing.
\end{abstract}
\maketitle

\section{Introduction}
Resonances are of utmost importance in diverse fields such as engineering and life sciences~\cite{Gammaitoni1998,Pisarchik2019}. In dynamical systems, when a nonlinear oscillator is strongly driven, it usually phase locks to the external drive~\cite{Thompson2002}. If one investigates the frequency of the oscillator as a function of the driving frequency, the resulting curve may consist of an infinite of steps with a fractal dimension between $0$ and $1$, which is known as the Devil's staircase~\cite{Bak1982}. In the context of quantum systems a natural question is: How do the fractional and integer resonances influence the dynamics of many-body systems under periodic drive? Currently, it is clear that understanding the nonequilibrium dynamics of a quantum many-body system poses challenges on moving beyond the standard framework of statistical mechanics \cite{Dziarmaga2010,RevModPhys.83.863,Eisert:2015aa,Mitra_Quench,Heyl_2018} and the efficient numerical simulation on classical computers \cite{Orus:2019aa,SCHOLLWOCK201196,10.21468/SciPostPhysLectNotes.8}. Advances in manipulating many-body systems allow us now to prepare exotic nonequilibrium states of matter using programmable quantum simulators \cite{PRXQuantum.2.017003} such as cold atoms \cite{RevModPhys.80.885,Cheneau:2012aa,Bernien:2017aa,Choi1547}, trapped ions \cite{Lanyon57,Blatt:2012aa,Zhang:2017ab,Zhang:2017ac}, and superconducting circuits \cite{Roushan1175,Ma:2019aa,PhysRevLett.123.050502,PhysRevLett.125.170503,PhysRevResearch.3.033043,Neill195}. In particular, periodically driven quantum systems \cite{Floquet1,Floquet2} are an exciting arena for discovering nonequilibrium states without static analog. Paradigmatic examples are discrete time crystals \cite{PhysRevA.91.033617,PhysRevLett.117.090402,PhysRevLett.118.030401,Sacha_2017,PhysRevLett.123.150601,TimeCrystals,Pizzi:2021we,PhysRevLett.127.140602,PhysRevB.104.094308,PhysRevResearch.3.L042023}, dynamical many-body freezing \cite{PhysRevB.90.174407,PhysRevB.82.172402}, and Floquet prethermalization \cite{PhysRevLett.115.256803,PhysRevB.95.014112,Abanin:2017uy,PhysRevLett.116.120401,KUWAHARA201696,PhysRevX.10.021044,PhysRevA.105.012418}.   

When a high-frequency driving (larger than any frequency scale of the undriven
system) acts upon a quantum system, the Floquet Hamiltonian $\hat{H}_F$ can be defined approximately using the Magnus expansion~\cite{Magnus1,Blanes_2010}. Driving a many-body system on resonance or off-resonance has significant consequences in the effective Hamiltonian that governs the dynamics \cite{PhysRevLett.116.125301}. This is particularly appealing in the Bose-Hubbard model (BHM) \cite{PhysRevB.40.546,Jaksch1998}. In the strong interaction limit \cite{torre2021statistical}, where the on-site repulsion dominates over the hopping, many-body resonances appear whenever $\Delta E=U[\pm(n_i-n_j)+1]=m\Omega$, with $U$ being the on-site repulsion, $\Omega$ the driving frequency, $m \in \mathbb{Z}$, while $n_i(n_j)$ is the occupation number at site $i(j)$. The upper (lower) sign means a hopping from site $j \to i$ ($i\to j$) respectively. In particular, a resonant high-frequency modulation of the hopping rate leads to an exponentially low heating rate, thus producing a prethermal regime \cite{PhysRevX.10.021044,torre2021statistical}. 

In this work, we demonstrate the emergence of a Floquet prethermal and localized quantum phase in a broad class of many-body Hamiltonians, when second order processes rule the many-body dynamics. This occurs if the condition $\pm m_{\Omega}=\pm(m_j-m_l)+1$ is satisfied, where $j$ and $l$ represent next-nearest neighbor sites, $m_{\Omega}=2\Omega/U$, and $m_{j}$ is a quantum number that labels the local states of lattice constituents, for example, it may represent the occupation number of bosonic systems or spin-1 states. If the maximal occupation number per site is limited to $m_j\le2$ for bosonic systems or $m_j\in\{-1,0,1\}$ for spin-1 lattice systems, the resonance condition is satisfied only if $m_{\Omega}=1$, which implies a fractional driving frequency $\Omega=U/2$. To demonstrate the dominating emergence of fractional resonances, we use the Magnus expansion of the effective Hamiltonian owing to the high-frequency driving acting upon many-body systems that exhibit reflection and U(1) symmetries. In a digital-analog quantum simulation scheme~\cite{DAQS}, our findings, put together with Floquet protocols of unitary gates \cite{PhysRevA.105.012418}, may provide alternative paths for discovering new phases of matter out of equilibrium. 
 \begin{figure*}[t]
\centering
\includegraphics[scale=0.50]{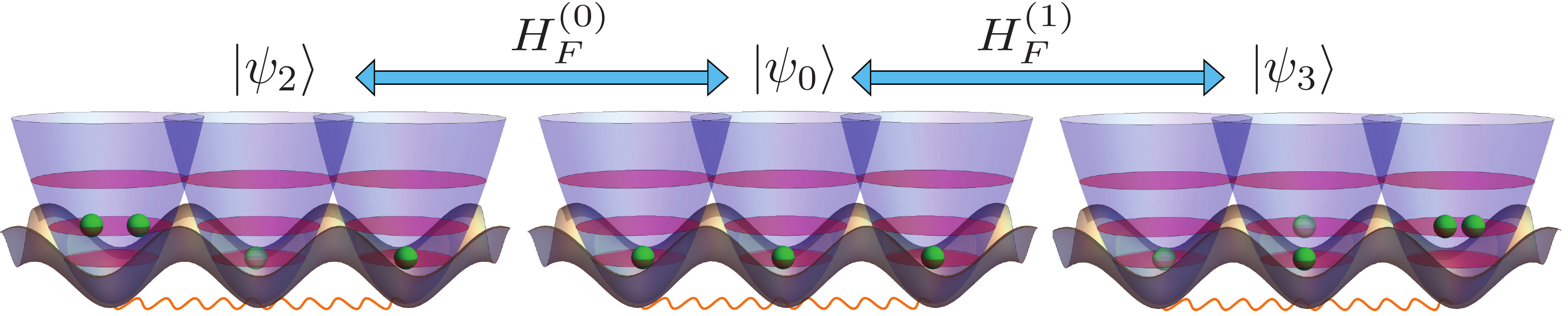}
\caption{The diagram represents many-body processes in the driven lattice model. The center panel represents the initial state with one excitation (full sphere) per site. The left panel represents single-excitation processes dominated by the frequency scale $J_0$, whereas the right panel virtual two-excitation processes dominated by $J_0^2/U$. The semitransparent sphere represents an empty state. The wiggle orange curve represents the modulated hopping rate $J(t)$.}
\label{Fig1}
\end{figure*} 

This article is organized as follows. In Sec.\ref{sec:II}, we present the generic many-body Hamiltonian exhibiting reflection and U(1) symmetry. The Hamiltonian is constructed using generic operators that follow a defined Lie algebra. In Sec.\ref{sec:III}, we discuss the emergence of integer and fractional many-body resonances using the Magnus expansion of the Floquet Hamiltonian. In particular, we demonstrate that at fractional resonances the leading term of the Magnus expansion is $\hat{H}_F^{(1)}$ with the frequency scale $J^{2}_0/U$ dominating the dynamics, where $J_0$ is the bare hopping rate. In contrast, at integer resonances, the zeroth order of the Magnus expansion, $\hat{H}_F^{(0)}$, plays the dominant role with a frequency scale $J_0$. These frequencies establish two scales for the spreading of excitations over the lattice. In Sec.\ref{sec:IV}, we exemplify our findings in a situation akin to experimental realization in superconducting circuits; namely, the three-site BHM initialized in a product state with unit filling, where we find analytical expressions for $\hat{H}_F^{(0)}$ and $\hat{H}_F^{(1)}$ and probabilities of accessible states at stroboscopic times. Then, in Sec.\ref{sec:V}, we extend our investigation to the many-body case that can be realized in diverse platforms \cite{Choi1547,Roushan1175,Ma:2019aa,PhysRevLett.123.050502,PhysRevLett.125.170503,PhysRevResearch.3.033043,Neill195}. We present numerical simulations of localization properties of quantum states \cite{WegnerF,Kramer_1993}, heating rate \cite{PhysRevB.104.134308,PhysRevLett.128.050604}, the half-chain von Neumann entropy \cite{PhysRevB.78.024410}, Loschmidt echo \cite{Heyl_2018}, and autocorrelation functions \cite{PhysRevResearch.3.033043} to quantify the critical slowing down characteristic of fractional resonances. Here, we also demonstrate the stability of the fractional resonance under perturbations in the resonance condition. In Sec.\ref{sec:VI} we discuss the emergence of fractional resonance in the $XXZ$ spin-1 model with anisotropy \cite{PhysRevB.67.104401,PhysRevLett.126.163203} and the Jaynes-Cummings-Hubbard model \cite{Greentree:2006aa,Hartmann:2006aa,PhysRevA.76.031805}. In Sec.\ref{sec:VII}, we present our concluding remarks. 

\section{The model}
\label{sec:II}
In quantum mechanics, we deal with symmetries related to groups that can be compact or not. A typical compact group is the special group of rotations in 3 dimensions SO(3). The Lorentz group turns out to be non-compact. A Lie algebra $\mathcal{G}$ can be interpreted as the tangent space of a Lie group at the identity element \cite{LieAG}. In this work, we focus on Lie algebras that admit a decomposition $\mathcal{G}=\mathcal{H}\bigoplus_\alpha \mathcal{G}_\alpha$, where $\mathcal{H}$ is the Cartan subalgebra. The ladder operators generate the subspaces $ \mathcal{G}_\alpha$.

To keep the discussion as general as possible, without losing the mathematical rigour, let us consider a Cartan algebra $\mathcal{H}_j$ for a subsystem at site $j$ with a single generator that we define as $\hat{O}_j$. Consequently, we consider local ladder operators $\hat{A}_j$ and $\hat{A}^{\dagger}_j$ such that they satisfy the algebraic relations $[\hat{O}_i,\hat{A}^{\dagger}_j]=\delta_{i,j}\hat{A}^{\dagger}_j$ and $[\hat{O}_i,\hat{A}_j]=-\delta_{i,j}\hat{A}_j$. Let us suppose that the local Hermitian operator $\hat{O}_j$ satisfies the eigenvalue equation $\hat{O}_j\ket{m_j}= m_j\ket{m_j}$, where $m_j$ is a quantum number that labels the local states, for example, it may represent the occupation number of bosonic systems or spin-1 states. Due to the algebraic structure, we have the relation $\hat{O}_j\hat{A}^{\dagger}_j\ket{m_j}=(\hat{A}^{\dagger}_j\hat{O}_j+[\hat{O}_j,\hat{A}^{\dagger}_j])\ket{m_j}=(m_j+1)\hat{A}^{\dagger}_j\ket{m_j}$. Using this algebra we can build a one-dimensional lattice with open boundary conditions, see Fig.~\ref{Fig1}, described by the generic Hamiltonian
\begin{align}
          \label{eq:HamAlgebra}
                  \hat{H}(t)=\hbar\sum^L_{j=1}\left(\omega \hat{O}_j+\frac{U}{2}\hat{O}^2_j\right)-\hbar J_0\cos{(\Omega t)}\sum^{L-1}_{j=1}(\hat{A}^{\dagger}_j\hat{A}_{j+1}+\hat{A}^{\dagger}_{j+1}\hat{A}_j)
\end{align}
The latter is composed of a local energy term $\hat{H}_0=\hbar\sum_{j=1}^L(\omega\hat{O}_j+U/2\hat{O}^2_j)$ that contains a single generator $\hat{O}_j$, and a time-dependent hopping term $\hat{H}_1(t)$ that represent the coupling between nearest neighboring sites via local ladder operators $\hat{A}_j$ and $\hat{A}^{\dagger}_j$. In the Hamiltonian (\ref{eq:HamAlgebra}) $\omega$, $U$, $J_0$, and $\Omega$ represent the local frequency scale, on-site interaction, bare hopping rate, and driving frequency, respectively. Since the Hamiltonian (\ref{eq:HamAlgebra}) exhibits U(1) symmetry, $e^{i\theta\hat{\mathcal{N}}}\hat{H}(t)e^{-i\theta\hat{\mathcal{N}}}=\hat{H}(t)$, where $\hat{\mathcal{N}}=\sum_{j=1}^L\hat{O}_j$, the term $\sum^N_{j=1}\omega \hat{O}_j$ is a constant of motion and does not play any role in the calculations. Also, since we consider open boundary conditions, the model exhibits parity symmetry such that $[\hat{H}(t),\hat{P}]=0$, where $\hat{P}\ket{m_1,m_2\hdots,m_L}=\ket{m_L,\hdots,m_2,m_1}$. Along this work we consider the strongly interacting regime characterized by $U/J_0\gg 1$ \cite{Cheneau:2012aa}, and within the subspace with filling factor $N/L=1$ for bosonic systems and total magnetization $\langle \hat{S}_z\rangle=0$ for spin-1 systems. For bosonic systems, the above condition will allow us to truncate the local Hilbert space to a maximum occupation number $n_{\rm max}=2$ when working with a large lattice size $L>6$.

In order to gain physical intuition on the processes that may occur due to hopping events, let us move to a rotating frame with respect to $\hat{H}_0$. The resulting Hamiltonian simply reads
\begin{equation}
\begin{aligned}
\tilde{H}_{I}(t)&=e^{\frac{i}{\hbar}\hat{H}_0t} \hat{H}(t)e^{-\frac{i}{\hbar}\hat{H}_0t}\\
&=-\hbar J_0\cos{(\Omega t)}\sum^{L-1}_{j=1}(e^{iU t(\hat{O}_{j+1}-\hat{O}_{j}- 1)}\hat{A}^{\dagger}_j\hat{A}_{j+1}\\
&+e^{-iU t(\hat{O}_{j+1}-\hat{O}_{j}+1)}\hat{A}^{\dagger}_{j+1}\hat{A}_j)
\ .
\label{HINT}
\end{aligned}
\end{equation} 
Notice that there are two characteristic frequencies in the Hamiltonian (\ref{HINT}), the driving frequency $\Omega$ and the on-site interaction $U$ which leads to a local anharmonic spectrum, see Appendix \ref{appendixA} for a detailed derivation of Eq.~(\ref{HINT}). 

The Hamiltonian~(\ref{HINT}) is not strictly periodic neither in $\Omega$ nor $U$; however, as we will prove in the next section, the Hamiltonian becomes periodic at fractional $\Omega=U/2$ and integer $\Omega=U$ driving frequencies. In this case, the Hamiltonian satisfies $\tilde{H}_{I}(t+T)=\tilde{H}_{I}(t)$ with period $T=2\pi/\Omega$, and we can apply the Floquet theory for time-periodic Hamiltonians \cite{GRIFONI1998229}. Surprisingly, the fractional frequency $\Omega=U/2$ is a resonance condition where second-order hopping processes become the dominating emergent mechanism displaying a generic slowing down of the many-body dynamics, prethermalization and localization simultaneously, as we will prove in next section.   

If the time-dependent Hamiltonian of a system is periodic $\hat{H}(t+T)=\hat{H}(t)$ with the period $T=2\pi/\Omega$, the whole dynamics is captured by the unitary time evolution operator $\hat{U}(t,t_0)=\hat{P}(t,t_0)e^{-\frac{i}{\hbar}\hat{H}_F(t-t_0)}$, where $\hat{P}(t,t_0)$ is the periodic kick operator and $\hat{H}_F$ is the time-independent Floquet Hamiltonian  \cite{GRIFONI1998229}. Obtaining a closed form of the Floquet Hamiltonian is not trivial, and particularly difficult for a quantum many-body system due to the exponential growth of the Hilbert space. When the driving frequency is much larger than all natural frequency scales of the undriven system, $\hat{H}_F$ can be approximated using the Magnus expansion (ME) $\hat{H}_F=\sum_{n=0}^{
\infty}\hat{H}_F^{(n)}$ \cite{Magnus1,Blanes_2010}.  The first two terms of the series read 
\begin{subequations}
\begin{align}
\hat{H}_F^{(0)}=&\frac{1}{T}\int_0^{T}dt \hat{H}(t)\\
\hat{H}_F^{(1)}=&\frac{1}{2!Ti}\int_0^{T}dt_1\int_0^{t_1}dt_2[\hat{H}(t_1),\hat{H}(t_2)].
\end{align} 
\label{HF}
\end{subequations}
We will use the expressions above to study the many-body quantum dynamics in detail. 

\section{Many-body resonances}
\label{sec:III}
In this section, we provide a detailed demonstration of integer and fractional resonances starting from the generic Hamiltonian (\ref{HINT}), and discuss about its periodicity under both resonance conditions. Also, we provide a general demonstration of the main result of our work, namely, the fractional resonance becomes the dominating emergent phenomena that rules the many-body dynamics. The latter is a consequence of the disappearance of the zeroth-order term $\hat{H}_F^{(0)}$ in the Magnus expansion, which produces the general mechanism of slowing down in many-body Hamiltonians that exhibit reflection and U(1) symmetries.   

\subsection{Integer resonance}
Many-body resonances can be identified when applying the Hamiltonian $\tilde{H}_I(t)$ to the quantum state $\ket{m_1,\hdots,m_j,m_{j+1},\hdots,m_L}$. As we stated in the previous section, $m_j$ is a quantum number that labels the local Hilbert space of lattice constituents. The result simply reads
\begin{widetext}
\begin{align}
\tilde{H}_I(t)\ket{m_1,\hdots,m_j,m_{j+1},\hdots,m_L}&=-\hbar J(t)\sum_{j=1}^{L-1}\big[e^{iU t(m_{j+1}-m_j+1)}\sqrt{m_j(m_{j+1}+1)}\ket{m_1,\hdots,m_j-1,m_{j+1}+1,\hdots,m_L}\nonumber\\
&+e^{iUt(m_{j}-m_{j+1}+1)}\sqrt{(m_j+1)m_{j+1}}\ket{m_1,\hdots,m_j+1,m_{j+1}-1,\hdots,m_L}\big].
\label{HME}
\end{align}
\end{widetext} 

The first term in Eq.~(\ref{HME}) represents a hopping event from site $j\to j+1$, whereas the second represents a hopping event from site $j+1\to j$. The hopping processes involve a change in the on-site interaction $\Delta E=U[\pm(m_j-m_{j+1})+1]$, where the upper (lower) sign means a hopping from site $j+1 \to j$ ($j\to j+1$) respectively. Many-body resonances appear whenever $\Delta E=U[\pm(m_j-m_{j+1})+1]=m\Omega$ \cite{torre2021statistical}, where $m\in \mathbb{Z}$. For instance, the lowest-order available resonance occurs for $m=\pm 1$, which results in the condition $\pm(m_j-m_{j+1})+1=\pm m_{\Omega}$, with $m_{\Omega}=\Omega/U$. The upper (lower) sign on the right-hand side of the resonance condition represents an increase (decrease) of energy respectively. If the maximal occupation number per site is limited to $m_j\le2$ for bosonic systems or $m_j\in\{-1,0,1\}$ for spin-1 lattice systems, the resonance condition can be satisfied if $m_{\Omega}=1$ which results in the integer resonance $\Omega=U$. The above conclusion is supported by the fact we consider the strongly interacting regime characterized by $U/J_0\gg 1$ \cite{Cheneau:2012aa}, and within the subspace with filling factor $N/L=1$ for bosonic systems and total magnetization $\langle \hat{S}_z\rangle=0$ for spin-1 systems. 

It is worth mentioning that Eq.~(\ref{HME}) may allow us to write down matrix elements that involve time-dependent functions of the type $f(t)=e^{\pm i\Omega t}e^{iU\bar{m}t}$ where $\bar{m}\in \mathbb{Z}$. If $\Omega=U/2$, then $f(t+T)=f(T)$ with period $T=4\pi/U$. Instead, if $\Omega=U$, then $f(t+T)=f(T)$ with period $T=2\pi/U$, thus the periodicity of the time-dependent Hamiltonian under integer and fractional resonances is valid for any lattice size $L$.
\begin{figure}
\centering
\includegraphics[scale=0.26]{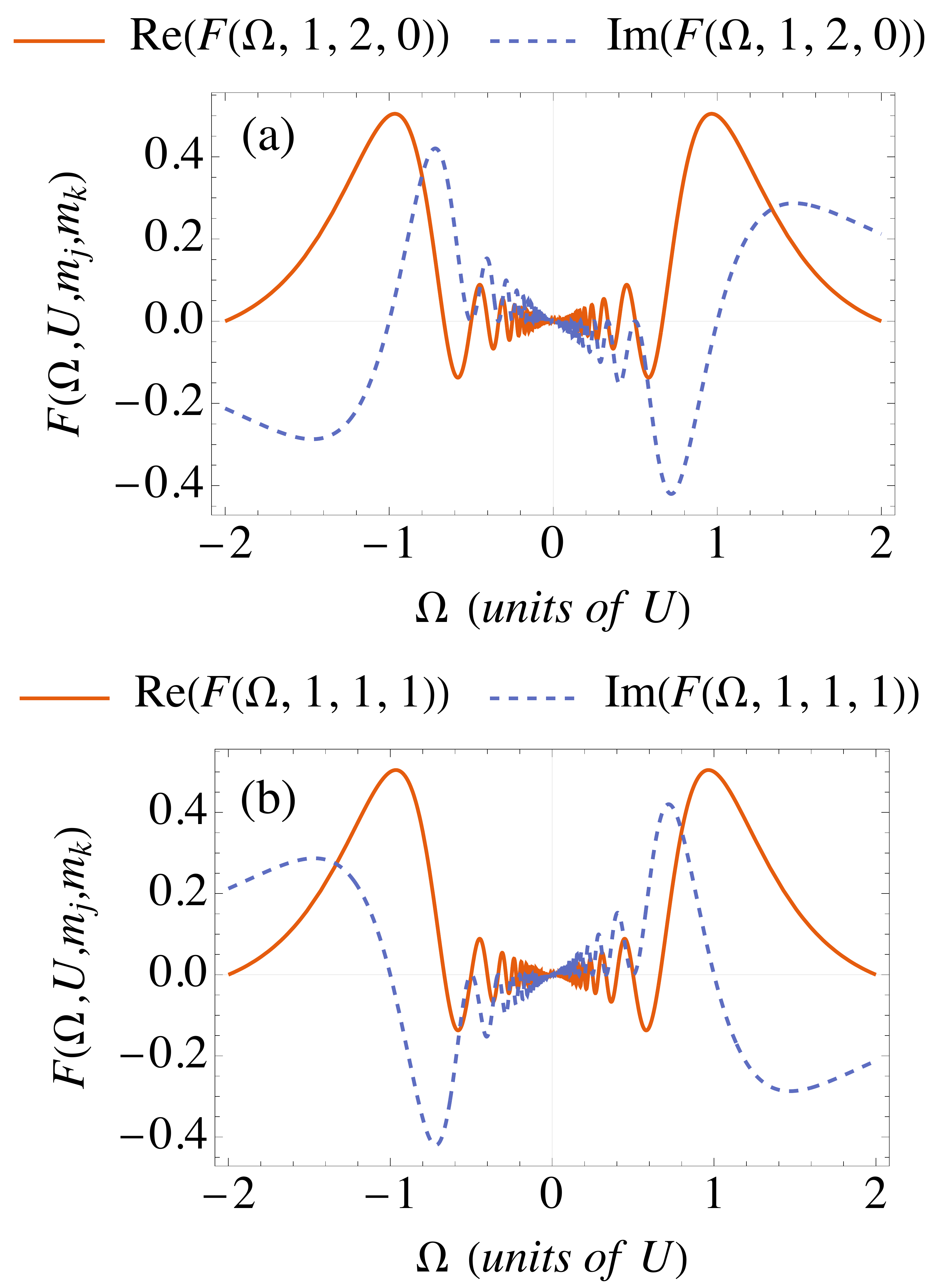}
\caption{(a,b) Real and imaginary parts of the function $F(\Omega,U,m_j,m_k)$ defined in Eq.~(\ref{matrixINT}) for a hopping processes from $j\to k$. In the upper (down) panel the configurations are $m_j=2$ and $m_k=0$ ($m_j=1$ and $m_k=1$) respectively. In both cases, the real part of $F(\Omega,U,m_j,m_k)$ is peaked near $\Omega=U$. In these plots we fixed $U=1$ for simplicity.}
\label{Fig2}
\end{figure}

Another useful way to derive the integer resonance is to consider the calculation of the time average of the matrix elements of Eq.~(\ref{HME}). The result is
\begin{equation}
\begin{aligned}
F(\Omega,U,n_j,n_k)&=\frac{1}{T}\int_0^T\cos(\Omega t)e^{iUt[\pm(m_{k}-m_j)+1]}\\
&=\frac{i U \Omega (m_k-m_j+1) \left(1 - e^{2 i \pi [\pm(m_k-m_j)+1]U/\Omega}\right)}{2 \pi  \left(U^2 [\pm(m_k-m_j)+1]^2-\Omega^2\right)}.
\label{matrixINT}
\end{aligned}
 \end{equation}
 
Figure~\ref{Fig2} we plot the function $F(\Omega,U,m_j,m_k)$ defined in Eq.~(\ref{matrixINT}) for hopping processes from $j\to k$, and for two different configurations, namely, $m_j=2, m_k=0$ (Fig.~\ref{Fig2}(a)) and $m_j=1, m_k=1$ (Fig.~\ref{Fig2}(b)) for bosonic systems or equivalently $m_j=1, m_k=-1$ and $m_j=0, m_k=0$ for spin-1 systems. In both cases, the real part of $F(\Omega,U,m_j,m_k)$ is peaked near $\Omega=U$, thus defining the integer resonance. Also, the integer resonance can be recognized by imposing the condition that Eq.~(\ref{matrixINT}) becomes an indeterminate form $0/0$, which occurs if $U [\pm(m_k-m_{j})+1]=\pm\Omega$, and we obtain the integer many-body resonance. Here, the time scale $J_0^{-1}$ dominates the system dynamics, with nearest-neighbor interactions playing the key role. This way of obtaining the integer resonance will be useful for recognizing the fractional resonance. 

An important discussion comes in order. Notice that the matrix elements $F(\Omega,U,m_j,m_k)=0$ if $(\pm M +1)U/\Omega=q$, where $q \in \mathbb{Z}\setminus\{0\}$ and we define $M=m_k-m_j\in\mathbb{Z}$. In other words, the contribution of the zeroth-order term in the ME (\ref{HF}) is exactly zero if the ratio $\Omega/U=(\pm M+1)/q$ is a rational number. The latter can be seen in Fig.~\ref{Fig2}(a,b) where $F(\Omega,U,m_j,m_k)$ exhibits zeros at values $\Omega/U=\pm 1/q$, where $q=\pm 2,\pm 3,\hdots$, and $q=\pm 2$ corresponding to the first zero. The disappearance of $\hat{H}_F^{(0)}$ allows us to conclude that higher-order terms in the ME~(\ref{HF}) dominate and lead to a generic slowing down of the many-body dynamics in models exhibiting U(1) and parity symmetry. The latter corresponds to the main result of our work. 

It is worth mentioning that our driving protocol is comparable to the one presented for the Fermi-Hubbard model (FHM) in Ref.~\cite{PhysRevLett.116.125301}. There, it has been proven that the FHM exhibits doublon association and dissociation processes when going on resonance, namely when the on-site interaction is an integer multiple of the driving frequency, where the leading order is $\hat{H}_F^{(0)}$. Also, in the off-resonance case where resonance effects can be ignored, the FHM exhibits an effective interacting spin model where the leading order is $\hat{H}_F^{(1)}$. In contrast, our driving protocol considers fractional resonance conditions, $\Omega/U=(\pm M+1)/q$, that lead to exactly zero contribution of $\hat{H}_F^{(0)}$. Therefore, our results provide a novel driving protocol of Floquet engineering where the effective Hamiltonian is not equal to the time-averaged Hamiltonian \cite{Floquet2}. Our results may provide new routes for quantum simulation using fractional resonances in Floquet engineering \cite{doi:10.1146/annurev-conmatphys-031218-013423,Weitenberg:2021ug}.  

\subsection{Fractional resonance}
In the previous subsection, we demonstrated that the zeroth-order term in the ME is exactly zero at fractional driving frequencies. Here, we focus on the first available fractional driving with $q=2$, and will demonstrate that processes that involve the virtual excitation of the middle site will be the leading contribution to the many-body dynamics. In particular, we will prove the fractional resonance condition $\pm m_{\Omega}=\pm(m_j-m_l)+1$, where $j$ and $l$ represent next-nearest neighbor sites and $m_{\Omega}=2\Omega/U$, by analyzing the commutator $[\hat{H}_I(t_1),\hat{H}_I(t_2)]$ in the ME (\ref{HF}). 

Analyzing the commutator $[\hat{H}_I(t_1),\hat{H}_I(t_2)]$, see Appendix~\ref{appendixB} for a detailed calculation, we recognize several hopping processes that may involve nearest-neighbor sites via operators $\hat{A}^{\dag}_j\hat{A}^{\dag}_j\hat{A}_{k}\hat{A}_{k}$, density-density interactions $\hat{A}^{\dag}_j\hat{A}_j\hat{A}^{\dag}_{k}\hat{A}_{k}$ or two excitation in the middle site $\hat{A}_{j}\hat{A}^{\dag}_k\hat{A}^{\dag}_k\hat{A}_{l}$, direct next-nearest neighbor sites $\hat{A}_j^{\dag}\hat{A}_{l}$, and virtual excitation of the middle site $\hat{A}_{j}\hat{A}^{\dag}_k\hat{A}_k\hat{A}^{\dag}_{l}$, where the indexes $j$, $k$, and $l$ represent left-most, middle, and right-most lattice sites.

Let us consider, for instance, the following term
\begin{align}
(e^{iUt_2}-1)e^{iU(\hat{O}_j-\hat{O}_{k}-1)t_1}e^{iU(\hat{O}_k-\hat{O}_{l}-1)t_2}\hat{A}^{\dag}_{j}\hat{A}_{k}\hat{A}^{\dag}_{k}\hat{A}_{l},
\label{OP1}
\end{align}
The above operator corresponds to the second term in the commutator (\ref{commutator}). Let us apply the operator (\ref{OP1}) to the generic state $\ket{m_1,\hdots,m_j,m_k,m_l,\hdots,m_L}$. The result involves the matrix element that connects the states $\ket{m_1,\hdots,m_j,m_k,m_l,\hdots,m_L}$ and $\ket{m_1,\hdots,m_j+1,m_k,m_l-1,\hdots,m_L}$, that is
\begin{widetext}  
\begin{align}
e^{iUt_1(m_j-m_k)}(e^{iUt_2}-1)e^{iUt_2(m_k-m_l)}(m_k+1)\sqrt{(m_j+1)m_l}\ket{m_1,\hdots,m_j+1,m_k,m_l-1,\hdots,m_L}.
\label{HF1}
\end{align}
\end{widetext}  
Now, let us compute the double integral 
\begin{align}
\frac{1}{2!Ti}\int_0^{T}dt_1\int_0^{t_1}dt_2J(t_1)J(t_2)e^{iUt_1(m_j-m_k)}(e^{iUt_2}-1)e^{iUt_2(m_k-m_l)},
\end{align}
in analogy with the calculation done in Eq.~(\ref{matrixINT}). The integral can be separated into two contributions as follows
\begin{widetext}
\begin{subequations}
\begin{align}
F_1(\Omega,U,m_j,m_k,m_l) &= \frac{1}{2!Ti}\int_0^{T}dt_1\int_0^{t_1}dt_2J(t_1)J(t_2)e^{iUt_1(m_j-m_k)}e^{iUt_2}e^{iUt_2(m_k-m_l)}\label{matrixFRAC1}\\
F_2(\Omega,U,m_j,m_k,m_l) &= \frac{1}{2!Ti}\int_0^{T}dt_1\int_0^{t_1}dt_2(t_1)J(t_2)e^{iUt_1(m_j-m_k)}e^{iUt_2(m_k-m_l)}.
\label{matrixFRAC2}
\end{align}
\end{subequations}
\end{widetext}

\begin{figure}[t]
\centering
\includegraphics[scale=0.22]{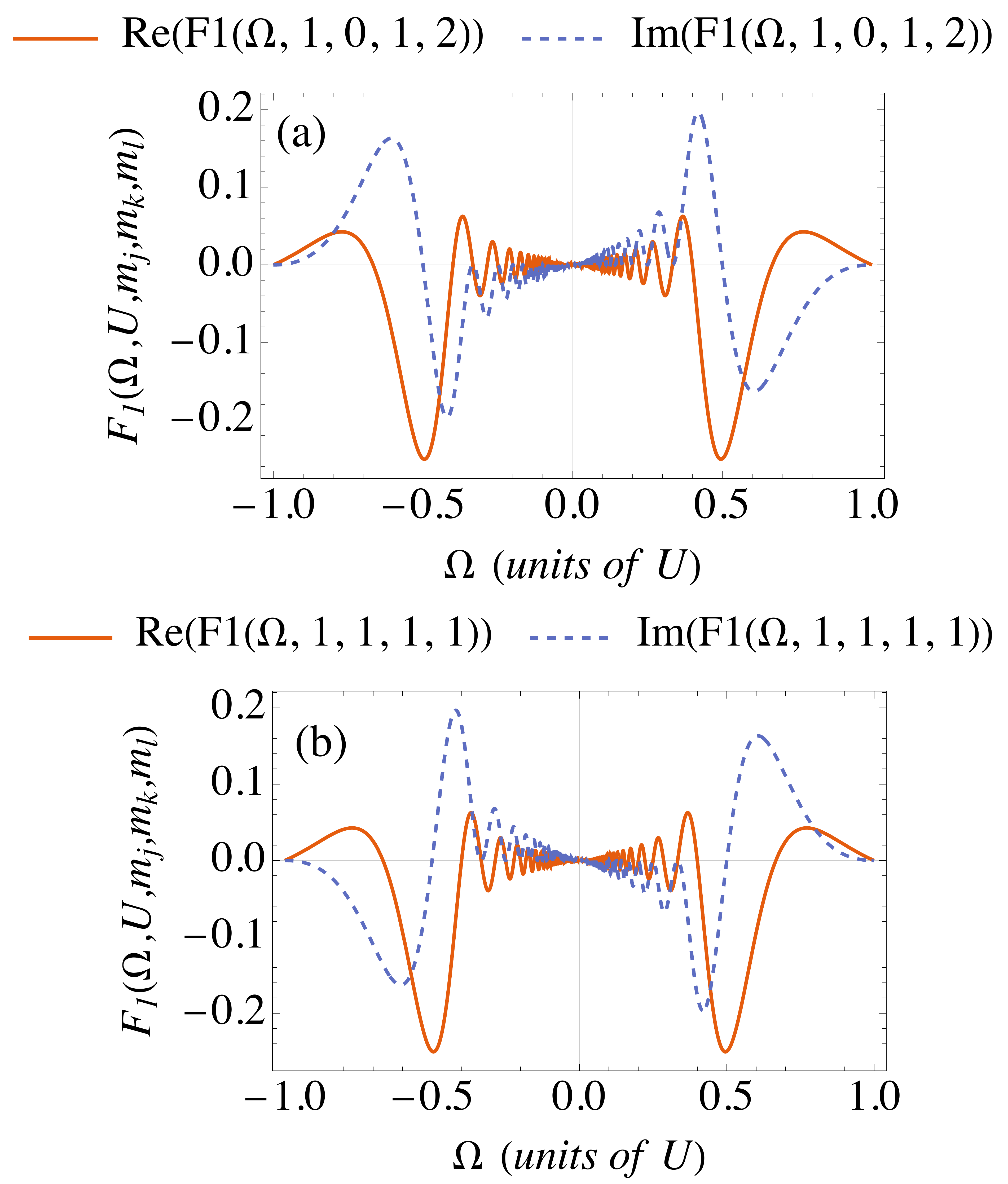}
\caption{(a,b) Real and imaginary parts of the function $F_1(\Omega,U,m_j,m_k,m_l)$ defined in Eq.~(\ref{matrixFRAC1}) for a hopping process from $l\to j$. In the upper (lower) panel the configurations are $m_j=0$, $m_k=1$ and $m_l=2$ ($m_j=1$, $m_k=1$ and $m_l=1$) respectively. In both cases, the real part of $F_1(\Omega,U,m_j,m_k,m_l)$ is peaked near $\Omega=U/2$. In these plots we fixed $U=1$ and for simplicity.}
\label{Fig3}
\end{figure}

In Fig.~\ref{Fig3} we plot the function $F_1(\Omega,U,m_j,m_k,m_l)$ defined in (\ref{matrixFRAC1}) for hopping process from $l\to j$, and for two different configurations, namely, $m_j=0$, $m_k=1$, $m_l=2$ (Fig.~\ref{Fig2}(a)) and  $m_j=1$, $m_k=1$, $m_l=1$ (Fig.~\ref{Fig2}(b)) for bosonic systems or equivalently $m_j=-1$, $m_k=0$, $m_l=1$ and $m_j=0$, $m_k=0$, $m_l=0$ for spin-1 systems. In both cases, the real part of $F_1(\Omega,U,m_j,m_k,m_l)$ is peaked near $\Omega=U/2$, thus defining the fractional resonance. It is worthwhile mentioning that the same analysis can be done for a hopping from $j\to l$. In analogy with the integer resonance, a detailed analysis of Eq.~(\ref{matrixFRAC1}) demonstrates that it becomes an indeterminate form $0/0$ if the system satisfies the condition 
\begin{align}
[\pm(m_j-m_{l})+1]=\pm m_\Omega,
\label{fractional}
\end{align} 
with $m_\Omega=2\Omega/U$. The upper (lower) sign on the right-hand side of Eq.~(\ref{fractional}) represents an increase (decrease) of energy respectively. If the maximal occupation number per site is limited to $m_j\le2$ for bosonic systems or $m_j\in\{-1,0,1\}$ for spin-1 lattice systems, the resonance condition is satisfied only if $m_{\Omega}=1$, which implies a fractional driving frequency $\Omega=U/2$.

We stress that at the fractional resonance $\Omega=U/2$ and using Mathematica \cite{Mathematica}, it can be shown that $\lim_{\Omega\to U/2} F_2(\Omega,U,m_j,m_k,m_l)\to 0$, but also other higher-order processes such as the creation of two particles/excitations at the intermediate site, so virtual excitations govern the system dynamics. Notice that density-density interactions such as $\hat{A}^{\dag}_j\hat{A}_j\hat{A}^{\dag}_k\hat{A}_k$, contribute only to the diagonal part of the effective Floquet Hamiltonian.

\section{Stroboscopic dynamics in the Bose-Hubbard Trimer}
\label{sec:IV}
Here, we present the quantum dynamics of a three-site lattice, see Fig.~\ref{Fig1}, in the high-frequency regime of a periodically modulated hopping scenario. We will discuss the integer ($\Omega=U$) and fractional ($\Omega=U/2$) driving and their effects on the system dynamics. As initial condition we consider a product state with one excitation per site for a fixed value $U/J_0=40$, that is, $|\psi(0)\rangle=\bigotimes_{j=1}^L\ket{1}_j$ where $L$ corresponds to the system size. Then, at $t=0$, we switch on the modulated hopping rate and let the system evolve under the Bose-Hubbard Hamiltonian
\begin{align}
\hat{H}(t) =  \hbar\sum_{j=1}^{L}(\omega \hat{a}^{\dag}_j\hat{a}_j+\frac{U}{2}\hat{a}^{\dag}_j\hat{a}^{\dag}_j\hat{a}_j\hat{a}_j)-\hbar J(t)\sum_{j=1}^{L-1}(\hat{a}^{\dag}_{j}\hat{a}_{j+1}+{\rm H.c}).   
\label{BHM}
\end{align}    

The number of states, here referred to as configurations, that may participate in the dynamics correspond to all possible configurations of $N$ particles distributed in $L$ lattice sites $D_N=(N+L-1)!/N!(L-1)!$. In the trimer case at unit filling $N/L=1$ there are $D_3=10$ configurations. The initial state $\ket{\psi_0}=\ket{111}$ has parity $p=+1$. Since the BHM preserves U(1) and parity symmetries, the dynamics will only involve states within the positive parity subspace  $\ket{\psi_0}$, $\ket{\psi_1}=\frac{1}{\sqrt{2}}(\ket{021} + \ket{120})$, $\ket{\psi_2}=\frac{1}{\sqrt{2}}(\ket{201} + \ket{102})$, $\ket{\psi_3}=\frac{1}{\sqrt{2}}(\ket{012} + \ket{210})$, $\ket{\psi_4}=\ket{030}$, $\ket{\psi_5}=\frac{1}{\sqrt{2}}(\ket{300} + \ket{003}$. In this basis the BHM, in the rotating frame with respect to $H_0=\hbar\sum_{j=1}^{L}(\omega \hat{a}^{\dag}_j\hat{a}_j+\frac{U}{2}\hat{a}^{\dag}_j\hat{a}^{\dag}_j\hat{a}_j\hat{a}_j)$, reads
\begin{equation}
\begin{aligned}
\hat{H}_{I}(t) =
   & - 2\hbar J_0\cos{(\Omega t)}|\psi_3\rangle\langle\psi_1| - \sqrt{6}\hbar J_0\cos{(\Omega t)}e^{2iUt} |\psi_4\rangle\langle\psi_1| \\ 
   &-\hbar J_0\cos{(\Omega t)}|\psi_3\rangle \langle\psi_2| - \sqrt{3}\hbar J_0\cos{(\Omega t)}e^{2iUt}|\psi_5\rangle\langle\psi_3|\\ 
    &-2\hbar J_0\cos{(\Omega t)}e^{iUt}(|\psi_1\rangle\langle\psi_0| + |\psi_2\rangle\langle\psi_0|)+ \rm{H.c.}
\label{HBH3}
\end{aligned}
\end{equation}

As we discussed in section \ref{sec:III}, the Hamiltonian (\ref{HBH3}) is not strictly periodic. However, it becomes periodic in the case of an integer $\Omega=U$ and fractional $\Omega=U/2$ resonances, and we can use the Floquet theory for describing the quantum dynamics.
\begin{figure}[t]
\centering
\includegraphics[scale=0.23]{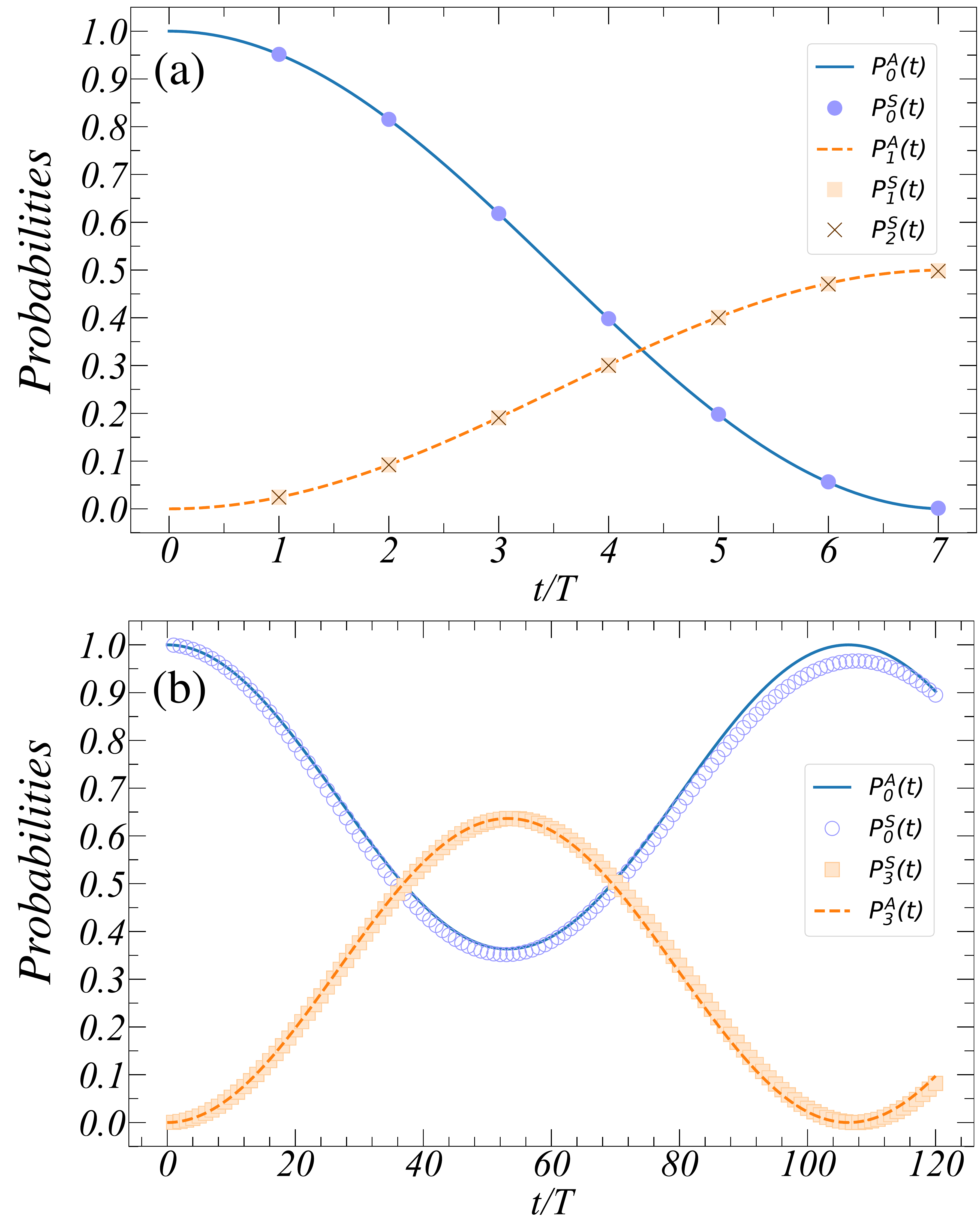}
\caption{Panels (a,b) plot the populations of the states $\ket{\psi_i}$ with $i=0,1,2,3$ for the integer $\Omega=U$ and fractional  $\Omega=U/2$ resonances, respectively. In (a,b) $P_0^{A}(t)$, $P_1^{A}(t)$, and $P_3^{A}(t)$ are populations analytically obtained from $\hat{H}_F^{(0)}$ and $\hat{H}_F^{(1)}$, while $P_i^{S}(t)$ stand for the stroboscopic dynamics. The initial state is $|\psi_0\rangle=\ket{111}$, and the parameters are $J_0=0.01\omega$, $U=40J_0$. We consider up to $n_{\rm max}=3$ particles per site with local Hilbert space dimension ${\rm dim}(\mathcal{H}_\ell)=4$.}
\label{Fig4}
\end{figure}

\subsection{Integer resonance}
Using the ME (\ref{HF}), the Floquet Hamiltonian to zeroth-order reads $\hat{H}_F^{(0)}=-\hbar J_0(\ket{\psi_0}\bra{\psi_1}+\ket{\psi_0}\bra{\psi_2}+{\rm H.c})$, whereas the matrix elements of $\hat{H}_F^{(1)}$ are of order $J^2_0/U$, so we neglect its contribution to the dynamics. Here, the time scale $J_0^{-1}$ dominates the system dynamics, with nearest-neighbor interactions playing the key role. The Schr\"{o}dinger equation can be solved by diagonalizing $\hat{H}_F^{(0)}$. Figure~\ref{Fig4}(a) shows the populations of states $|\psi_0\rangle$, $|\psi_1\rangle$ and $|\psi_2\rangle$ predicted from the Hamiltonian $\hat{H}_F^{(0)}$, and the stroboscopic evolution $\ket{\psi(nT)}=[\hat{U}(T,0)]^n\ket{\psi_0}$, where $\hat{U}(T,0)$ is the evolution operator in one period. The latter has been numerically computed from the Hamiltonian (\ref{BHM}) using exact diagonalization. We see a good agreement between the analytical (see Appendix \ref{appendixC}) and numerical predictions using the stroboscopic dynamics. Notice that after seven periods of the evolution, the initial population is fully transferred to states $\ket{\psi_1}$ and $\ket{\psi_2}$. Notice that the resonance condition $\Omega = U$ reduces the number of configurations that participate in the effective dynamics from $10$ to $5$.  
\begin{figure}[t]
\centering
\includegraphics[scale=0.24]{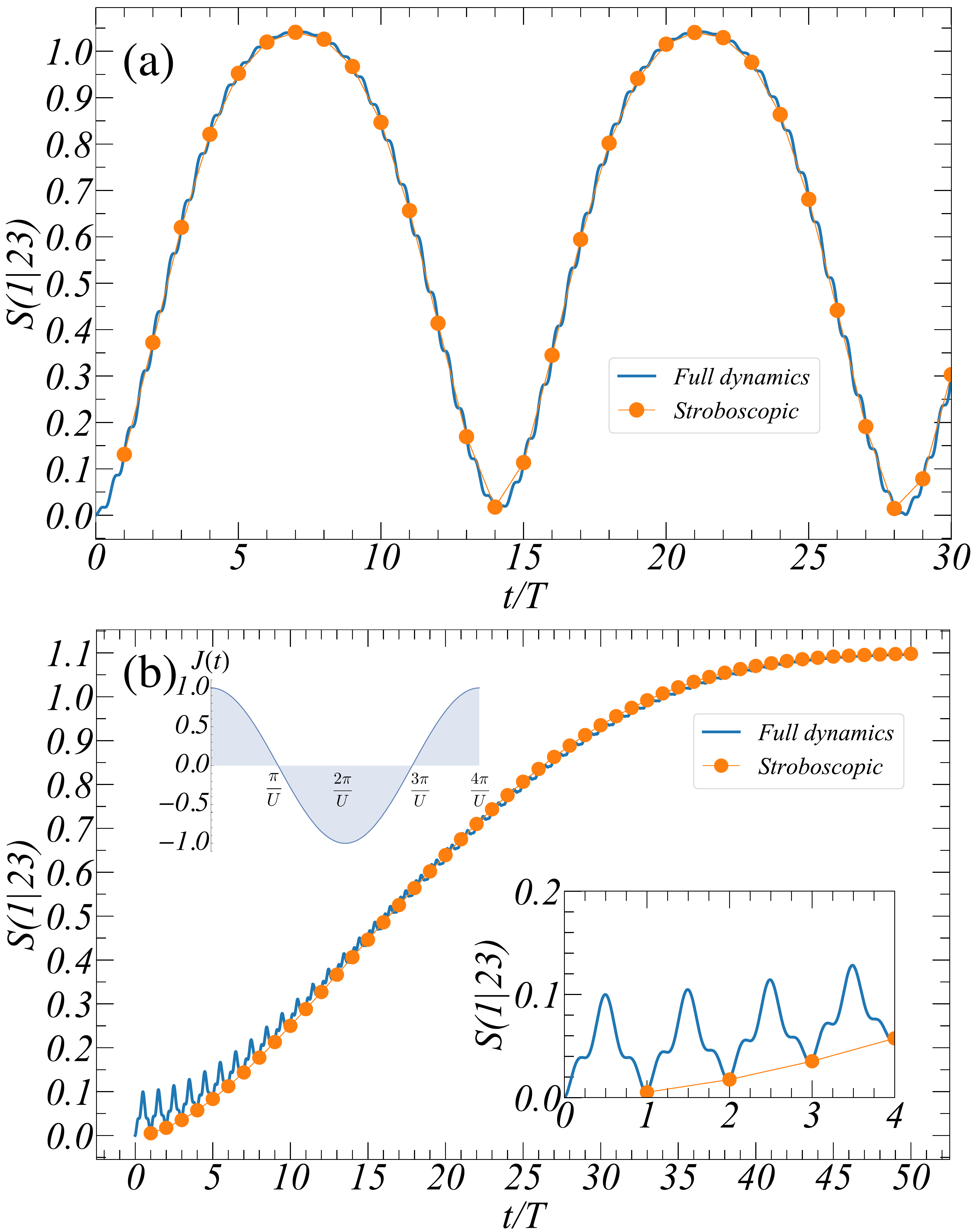}
\caption{Panels (a,b) show the von Neumann entropy of the bipartition $1|23$ as a function of time for integer $\Omega=U$ and fractional $\Omega=U/2$ resonances, respectively. We compare the full (solid lines) and stroboscopic (circles) dynamics. The lower inset shows the behavior of the von Neumann entropy after few periods of evolution. The initial state is $|\psi_0\rangle=\ket{111}$, and the parameters are $J_0=0.01\omega$, $U=40J_0$. We consider up to $n_{\rm max}=3$ particles per site with local Hilbert space dimension ${\rm dim}(\mathcal{H}_\ell)=4$.}
\label{Fig5}
\end{figure}

\subsection{Fractional resonance} 
At the fractional driving $\Omega=U/2$ and using the ME~(\ref{HF}), one can show that $\hat{H}_F^{(0)}=0$, whereas $\hat{H}_F^{(1)}$ reduces to a $2\times 2$ matrix in within the subspace $\{\ket{\psi_0},\ket{\psi_3}\}$, namely, 
$\hat{H}_F^{(1)}=\frac{16\hbar J^2_0}{3U}\ket{\psi_0}\bra{\psi_0}+\frac{4\hbar J^2_0}{5U}\ket{\psi_3}\bra{\psi_3}+\frac{3\hbar J^2_0}{U}(\ket{\psi_0}\bra{\psi_3}+\ket{\psi_3}\bra{\psi_0})$. Here, the frequency scale $3J_0^{2}/U$ rules the system dynamics, with next-nearest-neighbor interactions playing the key role. The latter results of the adiabatic elimination of states $\ket{\psi_1}$ and $\ket{\psi_2}$, thus producing a slower dynamics as compared with the integer resonance. The Schr\"{o}dinger equation can be solved by diagonalizing $\hat{H}_F^{(1)}$. Figure \ref{Fig4}(b) shows the populations of states $\ket{\psi_0}$ and $\ket{\psi_3}$ predicted from the Hamiltonian $\hat{H}_F^{(1)}$, and the stroboscopic evolution $\ket{\psi(nT)}=[\hat{U}(T,0)]^n\ket{\psi_0}$. We see a good agreement between the analytical (see Appendix~\ref{appendixC}) and numerical predictions using the stroboscopic dynamics. In contrast with the integer resonance, the initial population is not completely transferred to the state $\ket{\psi_3}$ and the state $\ket{\psi(t)}$ shows a strong overlap with the initial state. Also, the highest occupation probability of the state $\ket{\psi_3}$ occurs after $t\approx 50T$, which is a consequence of the slow dynamics. It is worth noticing that the fractional resonance $\Omega = U/2$ reduces further the number of configurations that participate in the effective dynamics from $10$ to $3$.

The previously exposed integer and fractional resonances have different time scales for spreading of bosonic particles with varying configurations. Their particular dynamical features also manifest in the bipartite entanglement dynamics. Figures~\ref{Fig5}(a,b) show the von Neumann entropy of the lattice bipartition $1|23$ as a function of time, where the upper (lower) panel stands for the integer (fractional) resonance. Notice the fast bipartite entanglement production in the integer resonance since the population of the initial state is completely transferred to states $\ket{\psi_1}$ and $\ket{\psi_2}$. The latter is opposite to the fractional resonance due to the slower dynamics and the substantial overlap between the state at time $t$ and the initial state (c.f. Fig.~\ref{Fig4}(b)).

\section{Many-body quantum dynamics}
\label{sec:V}
In this section, we quantify the effects of the fractional resonance $\Omega=U/2$ in the many-body context.
\subsection{Localization properties of the quantum states}
\label{localization}
Let us consider the BHM in finite lattice of $L=5$ sites initialized in the quantum state $|\psi(0)\rangle=\bigotimes_{j=1}^L\ket{1}_j$. At the unit filling condition, we work with a basis $\ket{l}$ of all possible $D_5=126$ configurations. In general, the time evolution of a given state can be written as a linear combination of the basis states $\ket{\psi(t)}=\sum_{l=1}^{D_N}c_l(t)\ket{l}$ with time dependent amplitudes $c_l(t)$. Now one may ask how many configurations “participate” in the dynamics. To measure this, we consider the participation ratio \cite{WegnerF,Kramer_1993}
\begin{align}
PR(t)=\frac{1}{\sum_{l=1}^{D_N}|c_l(t)|^4}.
\end{align} 
Since the dimension of the Hilbert space is $D_N$, when a state is fully delocalized the participation ratio is $PR(t)=D_N$. Besides, if the state is localized, only a single configuration participates in the superposition and $PR(t)=1$. 
\begin{figure}[t]
\centering
\includegraphics[scale=0.18]{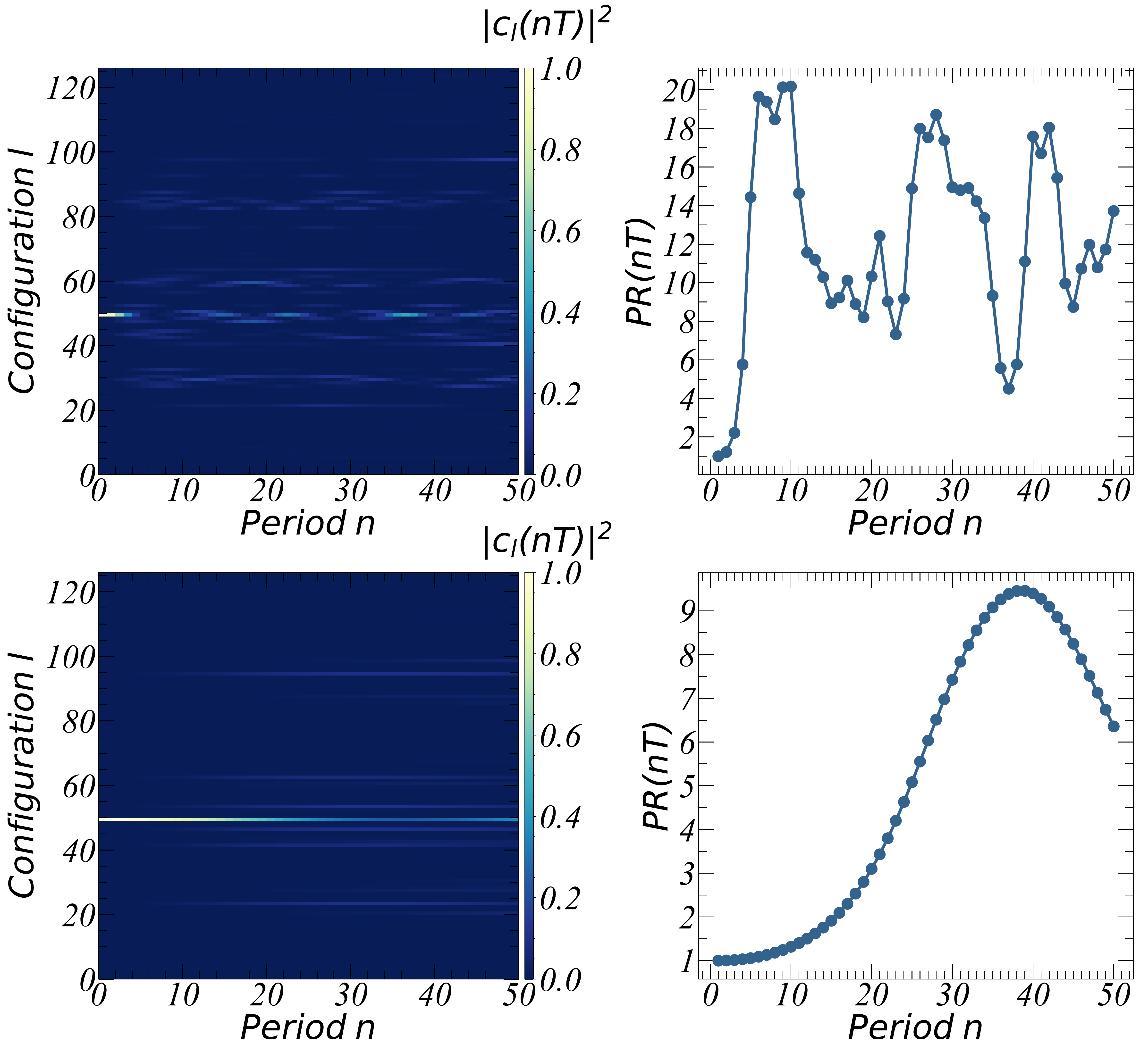}
\caption{Localization properties of quantum states. Here, we plot the distribution of populations associated to each configuration $|c_{l}(nT)|^2$ (left column) and the participation ratio $PR(nT)$ (right column), where $T=2\pi/\Omega$. The upper panel shows the results of the integer case $\Omega=U$, whereas the lower panel stands for the fractional case $\Omega=U/2$. In this simulation, we have numerically computed the effective Hamiltonian within the subspace with unit filling factor which contains $D_5=126$ configurations, starting from the Hamiltonian~(\ref{BHM}). The parameters are $J_0=0.01\omega$, $U=40J_0$, and we consider up to $n_{\rm max}=5$ particles per site with local Hilbert space dimension ${\rm dim}(\mathcal{H}_\ell)=6$.}
\label{Fig6}
\end{figure}

Figure~\ref{Fig6} shows the localization properties of quantum states at stroboscopic times by means of the population of the $l$th configuration $|c_l(nT)|^2$ (left column), and the participation ratio $PR(nT)$ (right column). For the integer resonance (upper panel), the system visits up to 20 configurations over time, unlike the fractional resonance (lower panel), where the system visits up to 10 configurations. Thus, the quantum state in the fractional resonance is more localized than in the integer resonance. We may also analyze the localization of quantum states by looking at the probability distribution $|c_l(nT)|^2$ of configurations $\ket{l}$ at stroboscopic times. The evolution is given by the repeated action of the unitary evolution operator in one period of the drive on the system's initial state $\ket{\psi(0)}$. As shown in the left column of Fig.~\ref{Fig6}, the probability distribution is more spread over a greater number of configurations in the integer resonance. In contrast, the fractional resonance regime is concentrated in a few configurations. In this context, the fractional resonance leads to less uncertainty in the quantum state than the integer resonance. 

\subsection{Heating Rate}
\begin{figure}[t]
\centering
\includegraphics[scale=0.21]{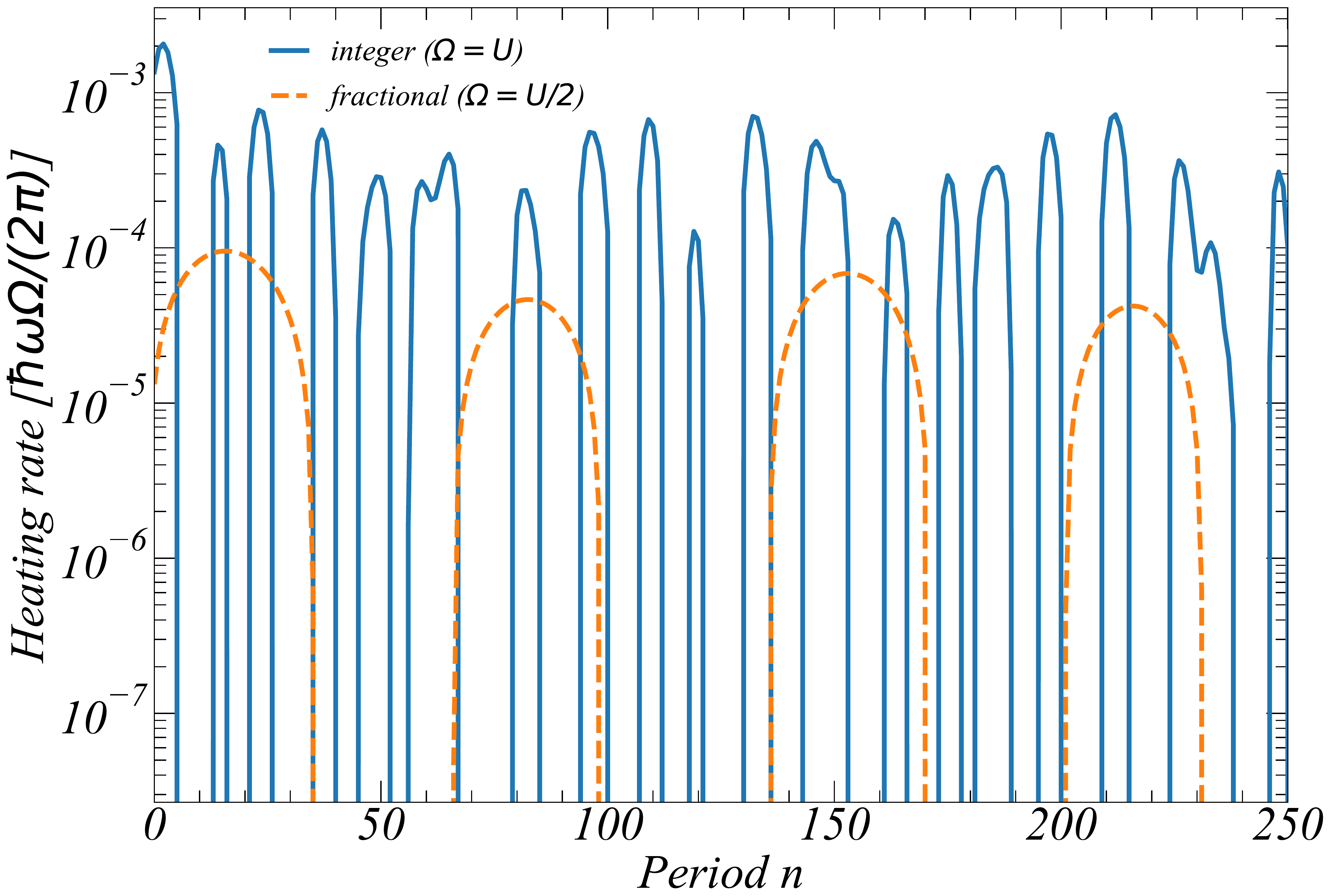}
\caption{The heating rate in semi-log scale as a function of the number of periods $n$. In this simulation we consider parameters $J_0=0.01\omega$, $U=40J_0$, $L=5$ sites, and we truncate the local Hilbert space up to $n_{\rm max}=5$ particles per site with local Hilbert space dimension ${\rm dim}(\mathcal{H}_\ell)=6$. }
\label{Fig7}
\end{figure}

The accurate calculation of the heating rate in our system is a subtle problem. The reason is that the system's response under the integer ($\Omega=U$) and fractional ($\Omega=U/2$) resonances is quite different. In the integer resonance case, analytical and numerical evidence show that the time-averaged Hamiltonian may describe the system dynamics with nearest-neighbor interactions playing the relevant role. In this case, one can use linear response theory for computing the heating rate \cite{torre2021statistical}. However, there is a dressing effect in the fractional resonance case. The Floquet Hamiltonian differs from the time-averaged Hamiltonian, leading to second-order processes dominating the system dynamics. Only recently, an accurate way of computing the heating rate in concrete systems, particularly with large-amplitude and high-frequency drivings, has been presented in Refs.~\cite{PhysRevB.104.134308,PhysRevLett.128.050604}. Here, we present the numerical calculation of the heating rate in a small lattice of $L=5$ sites. Intuitively, we expect the integer resonance to generate a larger heating rate than the fractional case since the latter generates a slower system response. In Fig.~\ref{Fig7}, we plot the heating rate calculated through the stroboscopic quantum evolution $\ket{\psi(nT)}=[\hat{U}(T,0)]^n\ket{\psi_0}$ with initial condition $\ket{\psi_0}$, and the average energy density $\epsilon_n=\bra{\psi(nT)}\hat{H}_0\ket{\psi(nT)}$ where $\hat{H}_0=\hbar\sum_{j=1}^L(\omega \hat{a}^{\dag}_j\hat{a}_j+\frac{U}{2}\hat{a}^\dag_j\hat{a}^\dag_j\hat{a}_j\hat{a}_j)$ is the undriven Hamiltonian. We compute the heating rate as the energy change in one period $(\epsilon_{n+1}-\epsilon_n)/T$ \cite{PhysRevB.104.134308}. 
 
We see the effect of many-body resonances on the heating rate. The fractional and integer resonances produce a low heating rate in the system, which signatures the prethermal regime of our system. Also, notice that the fractional resonance has a smoother heating rate than the integer resonance, reflecting the slower system's response under hopping driving. Here, we also identify an exciting problem of computing the heating rate for lattice sizes $L\ge 10$ that will be presented in a future work following the references \cite{PhysRevB.104.134308,PhysRevLett.128.050604}.  

\subsection{Half-chain von Neumann entropy, Loschmidt echo, and autocorrelation functions}
\begin{figure}[t]
\centering
\includegraphics[scale=0.108]{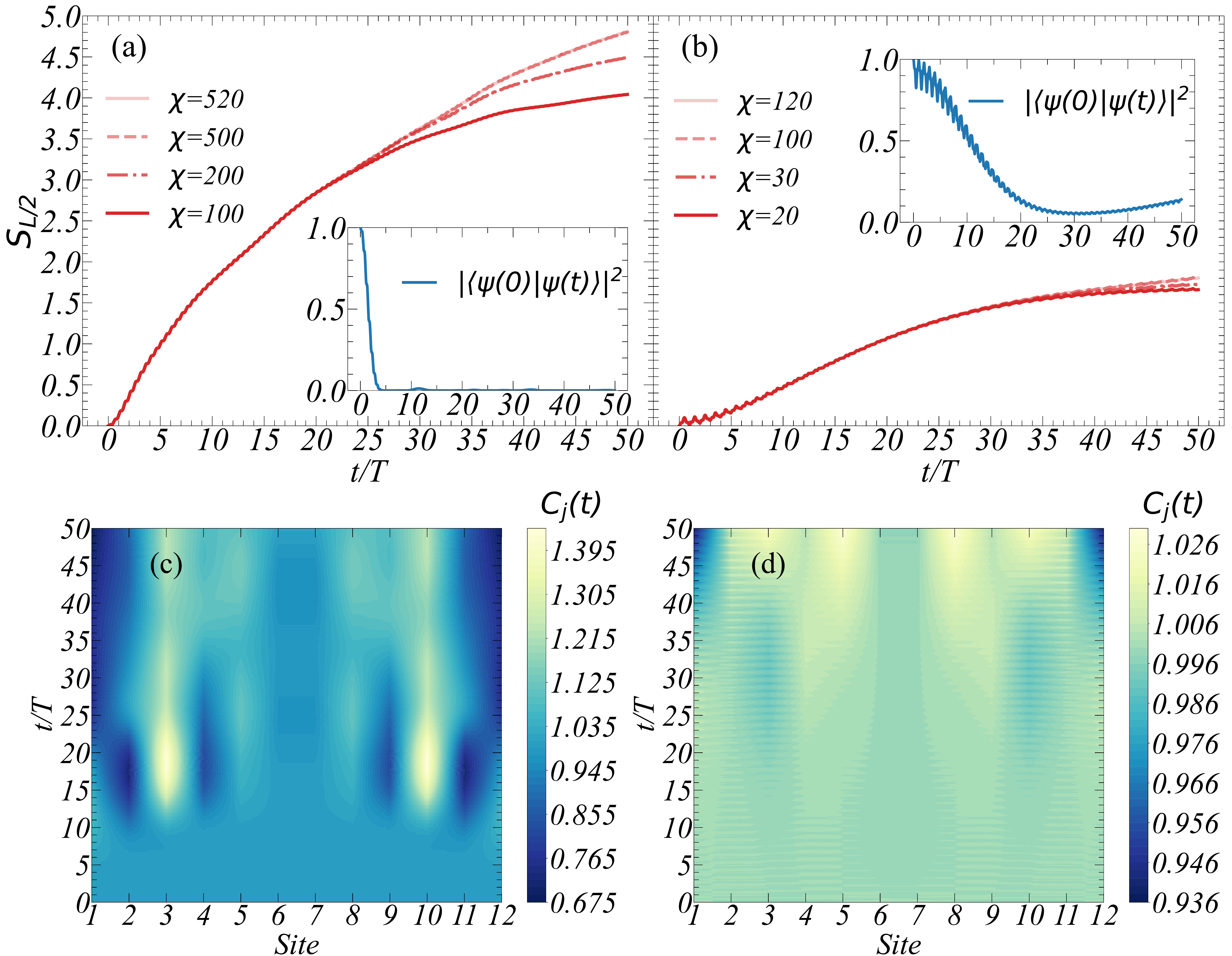}
\caption{Half-chain von Neumann entropy as a function of time for a lattice of $L=12$ sites.  
The panels (a,b) show the results for integer $\Omega=U$ and fractional $\Omega=U/2$ resonances, respectively. The insets show the Loschmidt echo $|\bra{\psi(0)}\psi(t)\rangle|^2$. Autocorrelation functions $\mathcal{C}_j(t)$ (see the main text) per site as a function of time are shown in panels (c,d) for integer $\Omega=U$ and fractional $\Omega=U/2$ resonances. For the integer resonance, autocorrelations $\mathcal{C}_j(t)$ experience fluctuations owing to the large uncertainty of the quantum state, as shown in panel (a). In the fractional resonance, fluctuations in $\mathcal{C}_j(t)$ are moderate owing to the large overlap between the initial state $\ket{\psi(0)}$ and the quantum state $\ket{\psi(t)}$, see the inset of panel (b). The parameters are $J_0=0.01\omega$, $U=40J_0$, and up to $n_{\rm max}=2$ particles per site.}
\label{Fig8}
\end{figure}
Let us now consider the nonequilibrium features of localization properties of quantum states characterized via the half-chain von Neumann entropy \cite{PhysRevB.78.024410}, Loschmidt echo \cite{Heyl_2018}, and autocorrelation functions \cite{PhysRevResearch.3.033043}. Due to two different time scales and physical processes, localization properties of quantum states in the integer and fractional resonances quantify how the information is spread over the Hilbert space. Suppose we want to extend the system size and make predictions for comparing with state-of-the-art quantum simulators. In that case, trustful numerical algorithms must consider information spreading and bipartite entanglement production. Here, we explore the scalability of the prethermalized and localized phase by considering a lattice of $L=12$ sites that has an immediate physical realization in superconducting circuits \cite{PhysRevLett.125.170503,PhysRevResearch.3.033043,Neill195}. At unit filling with the initial state $|\psi(0)\rangle=\bigotimes_{j=1}^L\ket{1}_j$, the Hilbert space configurations number is huge $D_{12}=1352078$, so, in order to study the nonequilibrium dynamics of the lattice, we use the time-evolving block decimation (TEBD) algorithm \cite{TEBD} implemented in TeNPy \cite{tenpy}. We consider the second-order Suzuki-Trotter decomposition of the unitary evolution operator and time step $dt=0.02\omega^{-1}$. We compute relevant quantities for describing the dynamics of each $10 dt$ time step. In the strongly interacting limit $U/J_0\gg 1$, we truncate the local Hilbert space to a maximum of $n_{\max}=2$ particles per site. It is worthwhile noticing that the case $n_{\rm max}=3$ provides the same results, see Appendix \ref{appendixD}. 

Figure~\ref{Fig8}(a,b) shows the half-chain von Neumann entropy for the integer (left panel) and fractional (right panel) resonances as a function of time. At the integer resonance, where first-order processes at a time scale $J_0^{-1}$ dominate the dynamics, the TEBD algorithm needs a large bond dimension $\chi=520$ to reach convergence within the simulating time, owing to the fast production of bipartite entanglement over the dynamics. In contrast, the TEBD algorithm needs a moderate bond dimension $\chi=120$ to reach convergence in the fractional resonance due to the slow production of bipartite entanglement. The resulting uncertainty of each regime is a consequence of the localization properties of the quantum states. We stress the truncation error $\epsilon_{\rm trunc}\le 10^{-8}$ at $t\approx50T$ in both resonance regimes, which implies the TEBD algorithm has an excellent performance within the simulating time. See Appendix \ref{appendixD} for a detailed discussion on the truncation error as one increases the lattice size.

The dynamical features of the half-chain von Neumann entropy may allow us to characterize each prethermal state obtained from the fractional and integer resonances. Also, these prethermal states may be characterized by the Loschmidt echo $L(t)=|\bra{\psi(0)}\psi(t)\rangle|^2$, see the insets of Fig.~\ref{Fig8}(a,b). At the integer resonance, the quantum state quickly departs from the initial condition. In contrast, the fractional resonance leads to a strong overlap with the initial state at short times. These dynamical features reflect in the dynamics of local observables such as autocorrelation functions $\mathcal{C}_j(t)=(2\langle n_j(t)\rangle-1)(2\langle n_j(0)\rangle-1)$ \cite{PhysRevResearch.3.033043}, where $\langle n_j\rangle$ stands for the average occupation number at the $j$th lattice site. Figs.~\ref{Fig8}(c,d) show the autocorrelations for each lattice site as a function of time. The left (right) panel represents the integer (fractional) resonances. The integer resonance exhibits more fluctuations of the local number of particles (excitations) than the fractional resonance owing to the larger uncertainty of the quantum many-body state. We state that autocorrelation functions that can be measured in superconducting circuit lattices \cite{PhysRevResearch.3.033043} may be considered a hallmark for identifying each prethermal state.  

\subsection{Stability of the fractional resonance under perturbations}
\begin{figure}[t]
\centering
\includegraphics[scale=0.29]{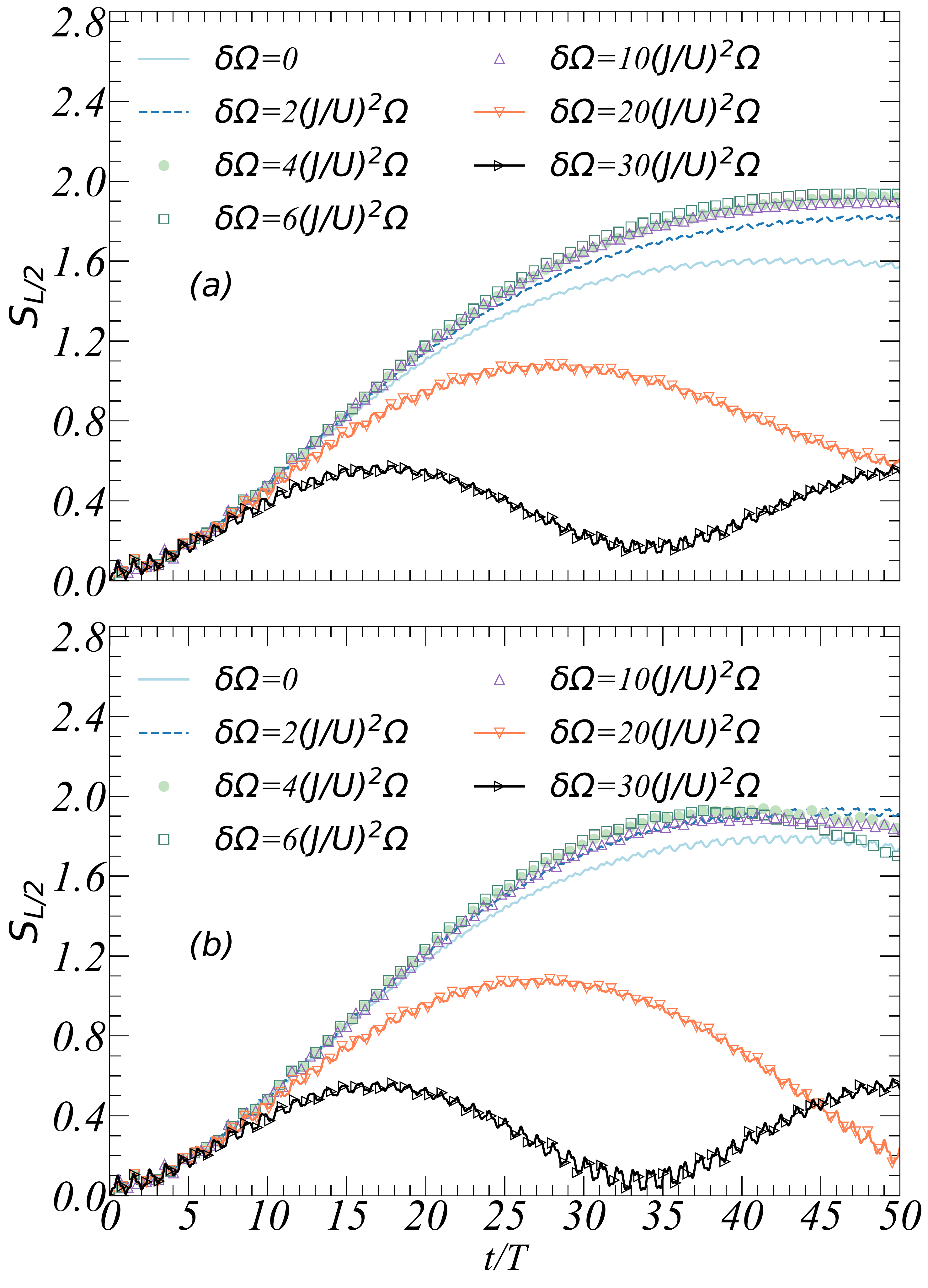}\\
\caption{The half-chain von Neumann entropy for a lattice size $L=4$ sites (a) and $L=12$ sites (b), and for various values of the perturbation $\delta\Omega/\Omega$. All curves have been calculated using parameters $J_0=0.01\omega$ and $U=40J_0$. In figure (a) we consider a maximum occupation number $n_{\rm max}=4$ per site. In figure (b), we truncate to a maximum occupation number $n_{\rm max}=2$ per site.}
\label{Fig9}
\end{figure}

Our work shows that fractional many-body resonances occur with an effective hopping rate $J_0^2/U$, which is slow compared to the integer many-body resonance. One may ask about the stability of the fractional resonance if the resonance frequency shifts $\Omega+\delta\Omega$ with $\delta\Omega/\Omega\gtrsim(J_0/U)^2$. We have performed numerical simulations for various values of $\delta\Omega/$. Figure~\ref{Fig9}(a,b) shows half-chain von Neumann entropy for a lattice of $L=4$ and $L=12$ sites respectively. The characteristic slowing down accompanying the fractional resonance is a stable phenomenon for perturbations $\delta\Omega/\Omega\le 10(J_0/U)^2$. However, we observe a strong suppression of the fractional resonance for values $\delta\Omega/\Omega>10(J_0/U)^2$, thus establishing a threshold value for the phenomenon's stability. We conclude that fractional resonance is a robust phenomenon for $\delta\Omega/\Omega\le 10(J_0/U)^2$ independently of the lattice size.  

\section{Universality of the fractional many-body resonance}
\label{sec:VI}
As we demonstrated in Sec.\ref{sec:III}, the fractional resonance and its characteristic slowing down of the many-body dynamics is not particular to the BHM. Instead, it can appear in several models that exhibit U(1) and parity symmetries such as the $XXZ$ spin-$1$ model with anisotropy \cite{PhysRevB.67.104401,PhysRevLett.126.163203}, the Jaynes-Cummings-Hubbard model \cite{Greentree:2006aa,Hartmann:2006aa,PhysRevA.76.031805}, and spin ladders (see Appendix \ref{appendixE}) as we will prove next
\subsection{Integer and fractional resonances in the $XXZ$ spin-1 model with anisotropy}
In this section, we prove the appearance of the integer and fractional resonance in the $XXZ$ spin-1 model with anisotropy described by the Hamiltonian \cite{PhysRevB.67.104401,PhysRevLett.126.163203}
\begin{align}
\hat{H} =\frac{\hbar U}{2}\sum_{j=1}^{L}(\hat{S}^z_j)^2 + \hbar J_0\cos(\Omega t)\sum_{j=1}^{L-1}(\hat{S}^{+}_j\hat{S}^{-}_{j+1}+\hat{S}^{+}_{j+1}\hat{S}^{-}_{j}),
\end{align} 
where $\hat{S}^z$, $\hat{S}^{\pm}$ are spin-$1$ operators that satisfy the commutation relations $[\hat{S}^z,\hat{S}^{\pm}]=\pm \hat{S}^{\pm}$, $[\hat{S}^+,\hat{S}^-]=\hat{S}^z$. This model also considers the competition between an anharmonic local spectrum and nearest-neighbor hopping. In order to prove the appearance of integer and fractional resonances, we consider the numerical simulation of a trimer with initial state with zero magnetization along the $z$ direction, namely, $\ket{\psi(0)}=\ket{0}\ket{0}\ket{0}$, and parameters $U=40J_0$. Notice that we use the local orthonormal basis for each spin as $\{\ket{m_j}\}$ with $m_j=-1,0,1$. Since the lattice exhibits reflection symmetry, and the total magnetization along the $z$ axis is conserved, the system will evolve within the positive parity sector whose states are
\begin{subequations}
\begin{align}
\ket{\psi_0}&=\ket{0}\ket{0}\ket{0}\nonumber\\
\ket{\psi_1}&=\frac{1}{\sqrt{2}}(\ket{1}\ket{-1}\ket{0}+\ket{0}\ket{-1}\ket{1})\nonumber\\
\ket{\psi_2}&=\frac{1}{\sqrt{2}}(\ket{-1}\ket{1}\ket{0}+\ket{0}\ket{1}\ket{-1})\nonumber\\
\ket{\psi_3}&=\frac{1}{\sqrt{2}}(\ket{1}\ket{0}\ket{-1}+\ket{-1}\ket{0}\ket{1})\nonumber.
\end{align}  
\end{subequations}

\begin{figure}[t]
\centering
\includegraphics[scale=0.27]{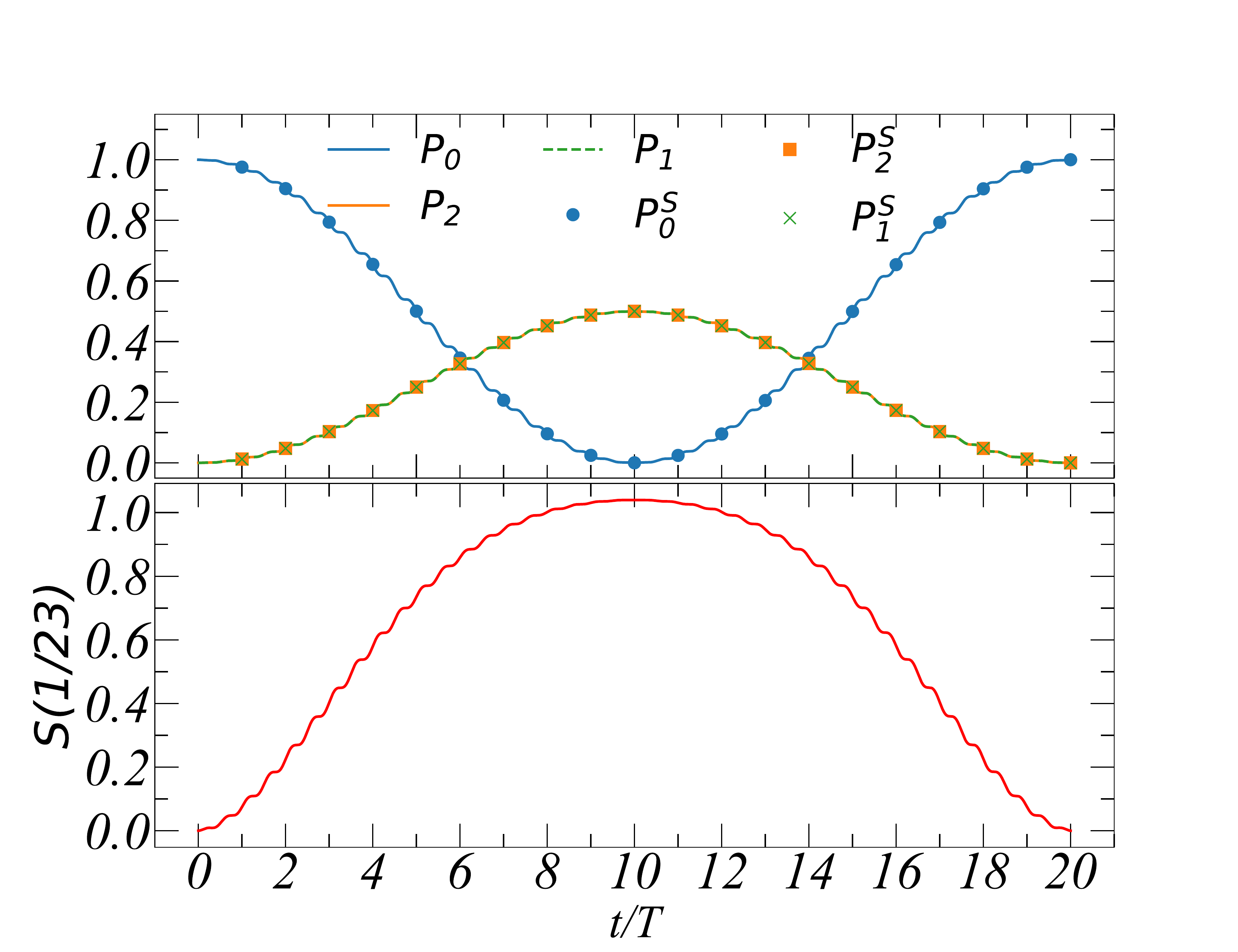}
\caption{Integer resonance case $\Omega=U$. The upper panel shows the populations of accesible states within the positive parity sector using the full and stroboscopic dynamics (markers). The lower panel shows the von Neumann entropy of the bipartition $1|23$.}
\label{Fig10}
\end{figure}

\begin{figure}[t]
\centering
\includegraphics[scale=0.27]{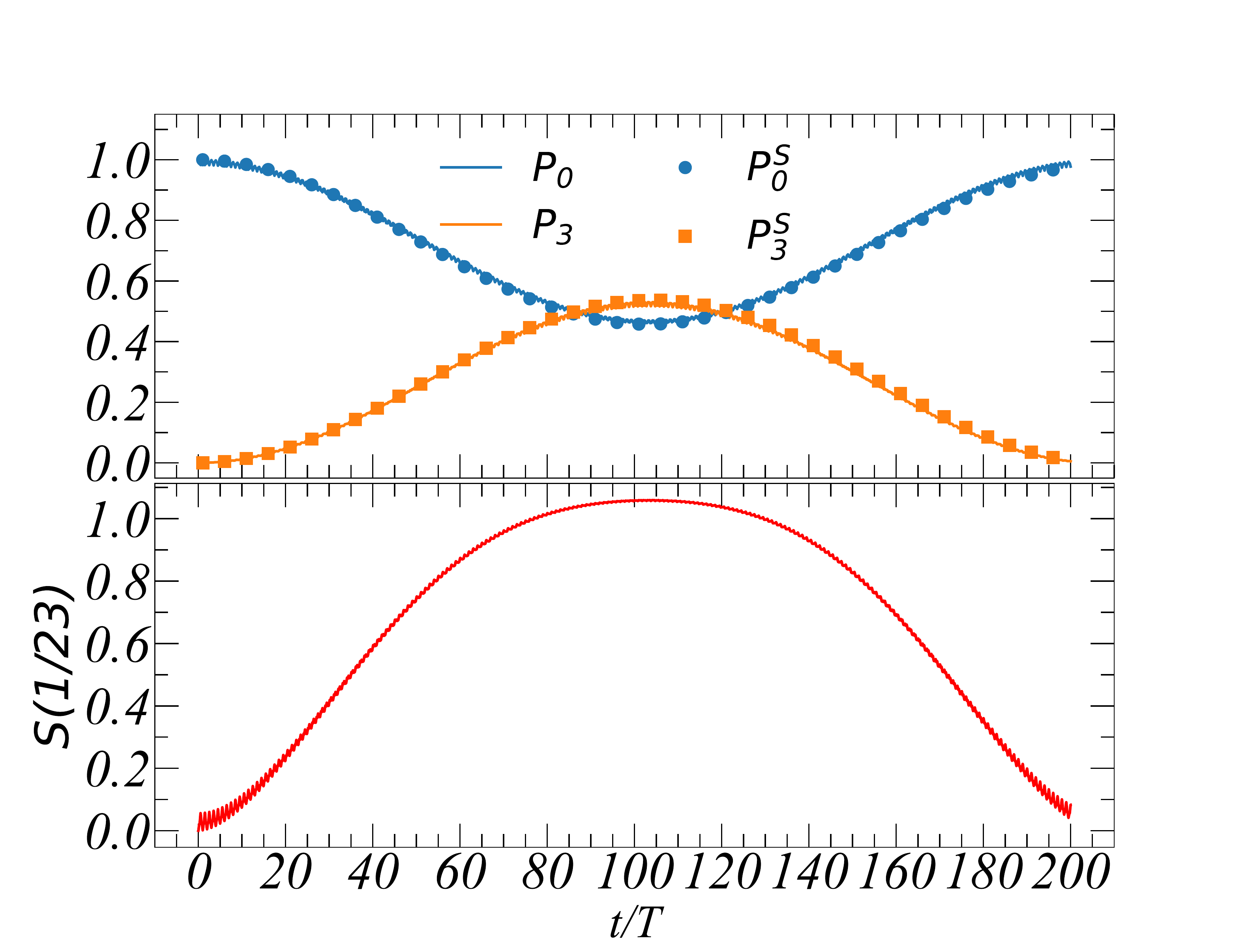}
\caption{Fractional resonance $\Omega=U/2$. The upper panel shows the populations of accessible states within the positive parity sector using the full and stroboscopic dynamics (markers). The lower panel shows the von Neumann entropy of the bipartition $1|23$.}
\label{Fig11}
\end{figure}

At the integer resonance condition $\Omega=U$, in Fig.~\ref{Fig10} we plot the probabilities of accessible states and the von Neumann entropy of the bipartition $1|23$. In analogy with the Bose-Hubbard model described previously, the integer resonance condition activates first-order processes where the initial population is transferred to states $\ket{\psi_1}$ and $\ket{\psi_2}$. In contrast, state $\ket{\psi_3}$ is not populated (not shown in the figure). Here, the populations are defined as $P_j(t)=|\langle \psi_j|\psi(t)\rangle |^2$. Markers correspond to the stroboscopic evolution. At the fractional resonance $\Omega=U/2$, in Fig.~\ref{Fig11} we plot the probabilities of accessible states and the von Neumann entropy of the bipartition $1|23$. The fractional resonance condition activates second-order processes where the initial population is transferred to state $\ket{\psi_3}$, whereas state $\ket{\psi_1}$ and $\ket{\psi_1}$ are not populated (not shown in the figure). Here, the populations are defined as $P_j(t)=|\langle \psi_j|\psi(t)\rangle |^2$. Markers correspond to the stroboscopic evolution. It is clear the slowing down accompanying the fractional resonance condition. 

\subsection{Integer and fractional resonances in the Jaynes-Cummings-Hubbard model}
Another Hamiltonian that exhibits integer and fractional resonances is the Jaynes-Cummings-Hubbard model \cite{Greentree:2006aa,Hartmann:2006aa,PhysRevA.76.031805}, which describes strongly interacting light-matter systems via the Hamiltonian
\begin{align}
\hat{H}_{\rm JCH} &= \hbar \sum_{j=1}^{L} \omega \hat{a}_j^{\dag}\hat{a}_j +\omega_0\hat{\sigma}_j^+\hat{\sigma}_j^-+ g(\hat{\sigma}_j^+\hat{a}_j+\hat{\sigma}_j^-\hat{a}_j^{\dag})_j\nonumber\\
&-\hbar J_0\cos(\Omega t)\sum_{j=1}^{L-1}(\hat{a}^{\dag}_j\hat{a}_{j+1}+\hat{a}_{j}\hat{a}^{\dag}_{j+1}),
\label{HJCH}
\end{align}
where $\hat{a}_j(\hat{a}_j^{\dag})$ is the annihilation (creation) bosonic operator at the $j$th lattice site, $\hat{\sigma}^{+}_j(\hat{\sigma}^{-}_j)$ is the raising (lowering) operator acting on the $j$th two-level system (TLS) eigenbasis $\{\ket{\downarrow}_j,\ket{\uparrow}_j\}$, and $\omega$, $\omega_0$, and $g$ are the resonator frequency, TLS frequency, and light-matter coupling strength, respectively. In order to prove the appearance of integer and fractional resonances, we consider the numerical simulation of a trimer whose initial state has one excitation per site, namely, $\ket{\psi(0)}=\ket{1,-}\ket{1,-}\ket{1,-}$, and parameters $g=40J_0$. Notice that the JCH Hamiltonian exhibits the competition between the local anharmonic spectrum provided by the Jaynes-Cummings interaction, and the photon-photon hopping interaction. Also, the local spectrum is described by hybrid light-matter states termed as polaritons defined by the upper ($+$) and lower $(-)$ polaritonic basis $\ket{n,\pm}_i=\gamma_{n\pm}\ket{\downarrow,n}_i+\rho_{n\pm}\ket{\uparrow,n-1}_i$ with energies $E^{\pm}_n=n\omega+\Delta/2\pm\chi(n)$. Here,
$\chi(n)=\sqrt{\Delta^2/4+g^2n}$, $\rho_{n+}=\cos(\theta_n/2)$,
$\gamma_{n+}=\sin(\theta_n/2)$, $\rho_{n-}=-\gamma_{n+}$,
$\gamma_{n-}=\rho_{n+}$, $\tan\theta_n=2g\sqrt{n}/\Delta$, and the detuning parameter $\Delta=\omega_0-\omega$. Also, one introduces the $j$th polaritonic
creation operators as $P^{\dag (n,\alpha)}_j=\ket{n,\alpha}_j\bra{0,-}$, where
$\alpha=\pm$ and we identify $\ket{0,-}\equiv\ket{\downarrow,0}$ and
$\ket{0,+}\equiv\ket{\emptyset}$ being a ket with all entries equal to
zero, that is, it represents an unphysical state. These
identifications imply $\gamma_{0-}=1$ and
$\gamma_{0+}=\rho_{0\pm}=0$. Using the above defined polaritonic basis, the
Hamiltonian (\ref{HJCH}) can be rewritten as
\cite{PhysRevA.76.031805,PhysRevA.80.023811,Pe_a_2020,PhysRevA.103.053708}.
\begin{widetext}
\begin{eqnarray}
\hat{H} =\hbar \sum^L_{j=1}\sum^{\infty}_{n=1}\sum_{\alpha=\pm}E_n^{\alpha} 
\hat{P}^{\dag (n,\alpha)}_j\hat{P}^{(n,\alpha)}_j\hat{P}^{\dag (n,\alpha)}_j\hat{P}^{(n,\alpha)}_j
-\hbar J(t)\sum^{L-1}_{j=1}\Big[\sum^{\infty}_{n,m=1}\sum_{\alpha,\alpha',\beta,\beta'=\pm}t_{n}^{\alpha\alpha'}
t_{m}^{\beta\beta'}\hat{P}^{\dag (n-1,\alpha)}_j\hat{P}^{(n,\alpha')}_j\hat{P}^{\dag (m,\beta)}_{j+1}\hat{P}^{(m-1,\beta')}_{j+1}+ {\rm H.c}\Big],\nonumber\\
\label{HPolariton}
\end{eqnarray}
\end{widetext}
where the matrix elements $t_{n}^{\alpha\alpha'}=\sqrt{n}\gamma_{(n-1)\alpha}\gamma_{n\alpha'}+\sqrt{n-1}\rho_{(n-1)\alpha}\rho_{n\alpha'}$. The first term in Eq.~(\ref{HPolariton}) stands for the local
polaritonic energy with an anharmonic spectrum and gives rise to an
effective on-site polaritonic repulsion. This is analog to the on-site photon repulsion in the Bose-Hubbard model~\cite{PhysRevB.40.546,Jaksch1998}. The last term in
Eq.~(\ref{HPolariton}) represents the polariton hopping between
resonators. The JCH Hamiltonian preserves the total number excitations described by the operator $\mathcal{\hat{N}}=\sum_{j=1}^L(\hat{a}^{\dag}_j\hat{a}_j+\hat{\sigma}^{+}_j\hat{\sigma}^{-}_j)$, and it exhibits reflection symmetry. Also, we work in the regime $g>4J_0$ which allows us neglecting the interchange of the upper and lower polaritonic branches \cite{PhysRevA.80.023811,Pe_a_2020,PhysRevA.103.053708}. In this case, the system will evolve within the positive parity sector whose states are
\begin{subequations}
\begin{align}
\ket{\psi_0}&=\ket{1,-}\ket{1,-}\ket{1,-}\nonumber\\
\ket{\psi_1}&=\frac{1}{\sqrt{2}}(\ket{2,-}\ket{0,-}\ket{1,-}+\ket{1,-}\ket{0,-}\ket{2,-})\nonumber\\
\ket{\psi_2}&=\frac{1}{\sqrt{2}}(\ket{0,-}\ket{2,-}\ket{1,-}+\ket{1,-}\ket{2,-}\ket{0,-})\nonumber\\
\ket{\psi_3}&=\frac{1}{\sqrt{2}}(\ket{2,-}\ket{1,-}\ket{0,-}+\ket{0,-}\ket{1,-}\ket{2,-})\nonumber.
\end{align}  
\end{subequations}

\begin{figure}[t]
\centering
\includegraphics[scale=0.27]{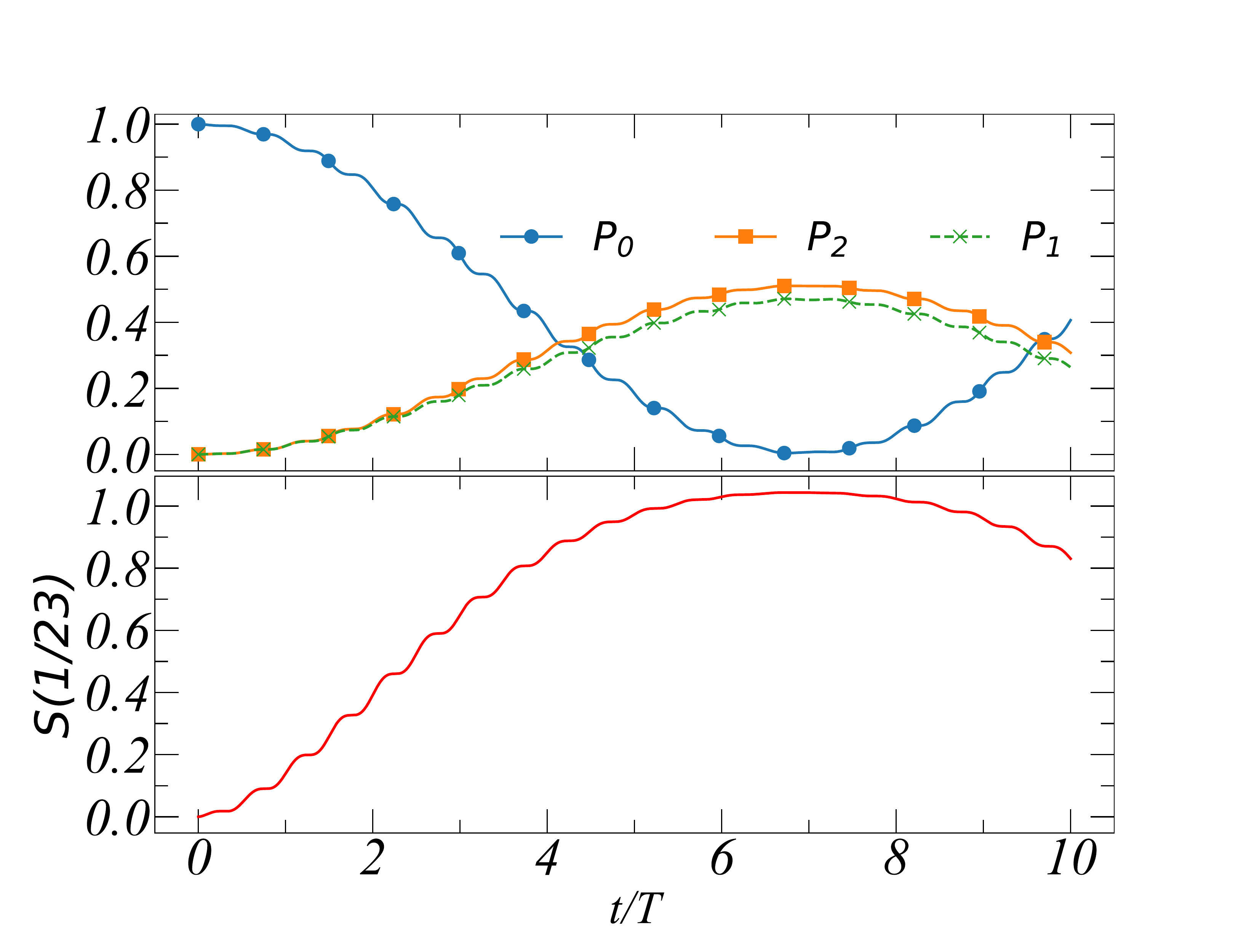}
\caption{Integer resonance case $\Omega=(2-\sqrt{2})g$. The upper panel shows the populations of accessible states within the positive parity sector using the full and stroboscopic dynamics. The lower panel shows the von Neumann entropy of the bipartition $1|23$.}
\label{Fig12}
\end{figure}
\begin{figure}[t]
\centering
\includegraphics[scale=0.27]{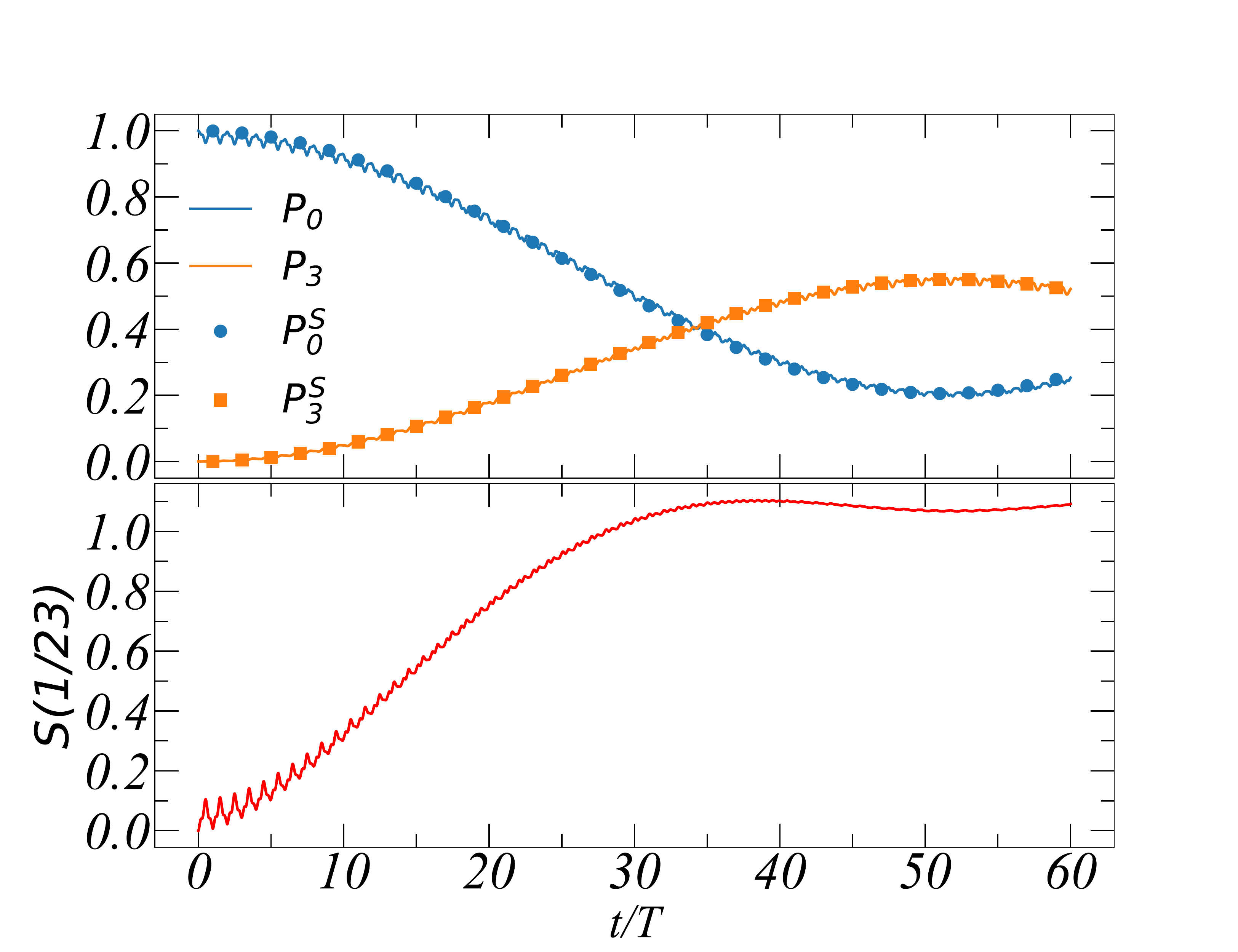}
\caption{Fractional resonance case $\Omega=(2-\sqrt{2})g/2$. The upper panel shows the populations of accessible states within the positive parity sector using the full and stroboscopic dynamics. The lower panel shows the von Neumann entropy of the bipartition $1|23$.}
\label{Fig13}
\end{figure}

Here we consider $\Delta=\omega_0-\omega=0$, which provides the largest anharmonic local spectrum. In this case, the integer many-body resonance condition reads $\Omega=(2-\sqrt{2})g$. In Fig.~\ref{Fig12}, we plot the probabilities of accessible states and the von Neumann entropy of the bipartition $1|23$. The integer resonance condition activates first-order processes where the initial population is transferred to states $\ket{\psi_1}$ and $\ket{\psi_2}$. In contrast, state $\ket{\psi_3}$ is negligibly populated (not shown in the figure). Nonetheless, the leakage from states $\ket{\psi_1}$ and $\ket{\psi_2}$ to the state $\ket{\psi_3}$ is not symmetric, which explains the asymmetry in the populations $P_1$ and $P_2$. Here, the populations are defined as $P_j(t)=|\langle \psi_j|\psi(t)\rangle |^2$. Markers correspond to the stroboscopic evolution.  

At the fractional resonance $\Omega=(2-\sqrt{2})g/2$, in Fig.~\ref{Fig13} we plot the probabilities of accessible states and the von Neumann entropy of the bipartition $1|23$. In analogy with the previous models, the fractional resonance condition activates second-order processes where the initial population is transferred to state $\ket{\psi_3}$, whereas state $\ket{\psi_1}$ and $\ket{\psi_2}$ are not populated (not shown in the figure). Here, the populations are defined as $P_j(t)=|\langle \psi_j|\psi(t)\rangle |^2$. Markers correspond to the stroboscopic evolution. Again, it is clear the slowing down accompanying the fractional resonance condition. 
\section{Conclusions}
\label{sec:VII}
We have provided robust evidence of the universality of our results by generalizing the fractional resonance to a broad class of many-body systems, all displaying more robust localization and slower heating rate compared with the integer resonance. We demonstrated the disappearance of the zeroth-order term in the Magnus expansion for fractional frequency drivings, so that higher-order terms become the leading contribution and explain the slowing down of the many-body dynamics in models that exhibit U(1) and parity symmetries. Also, the effect of the fractional resonance in many-body dynamics has been quantified through the von Neumann entropy, Loschmidt echo, and heating rate, thus proving clear evidence of less entanglement creation, more localized quantum states, and the slower system’s response in contrast at the integer resonances. In this way, fractional resonances and their dynamical features emerge as a collective phenomenon independent of the microscopic nature of each model, and present themselves as a general physical principle that can be used to develop quantum memories \cite{doi:10.1073/pnas.1819316116} for quantum technologies, provide a new route of quantum simulation with Floquet engineering where the higher-order terms dominates, the discovery of new phases of matter in periodically driven systems, a deeper understanding of the prethermal regime beyond the linear response \cite{PhysRevB.104.134308,PhysRevLett.128.050604}, and applications in quantum sensing \cite{PhysRevLett.127.080504}.

\section*{Acknowledgments}
We thank D. Rossini for useful discussions. R.P. acknowledges the support from Vicerrector\'ia de Postgrado USACH, F. T. acknowledges financial support from Fondo Nacional de Investigaciones Científicas
y Tecnológicas (FONDECYT, Chile) under grants 1211902 and Centro de Nanociencia y Nanotecnología CEDENNA,
Financiamiento Basal para Centros Científicos y Tecnológicos de Excelencia AFB180001, G.R acnowledges the support from FONDECYT grant No. 1190727.

\appendix

\section{Generic Hamiltonian in the interaction picture}
\label{appendixA}
Here, we show how to transform the generic Hamiltonian (\ref{eq:HamAlgebra}) into the interaction picture. Here, we find expressions like
\begin{align}
          \label{eq:OperatorEv}
                  \hat{A}^{\dagger}_i(t)=e^{\frac{i}{\hbar}\hat{H}_0t}\hat{A}^{\dagger}_ie^{-\frac{i}{\hbar}\hat{H}_0t}
                  \ .
\end{align}

By taking the derivative of this expression, we end up with terms like
\begin{align}
          \label{eq:OperatorEv}
                  \frac{d \hat{A}^{\dagger}_j(t)}{dt}=\frac{i U}{2}[\hat{O}^2_j(t),\hat{A}^{\dagger}_j(t)]=\frac{i U}{2}(2\hat{O}_j(t)-1)\hat{A}^{\dagger}_j(t)
                  \ .
\end{align}

From this we can get the time evolution $\hat{A}^{\dagger}_j(t)=\exp[\frac{iU t}{2}(2\hat{O}_j-1)]\hat{A}^{\dagger}_j$. Similarly, we can proof that $\hat{A}_j(t)=\exp[-\frac{iU t}{2}(2\hat{O}_j+1)]\hat{A}_j$. By using this, we obtain the explicit form for the Hamiltonian in the rotating frame [c.f. Eq.~(\ref{HINT})]
\begin{align}
\tilde{H}_{I}(t)&=e^{\frac{i}{\hbar}\hat{H}_0t} \hat{H}(t)e^{-\frac{i}{\hbar}\hat{H}_0t}\nonumber\\
&=-\hbar J_0\cos{(\Omega t)}\sum^{L-1}_{j=1}(e^{iU t(\hat{O}_{j+1}-\hat{O}_{j}- 1)}\hat{A}^{\dagger}_j\hat{A}_{j+1}\nonumber\\
&+e^{-iU t(\hat{O}_{j+1}-\hat{O}_{j}+1)}\hat{A}^{\dagger}_{j+1}\hat{A}_j).
\end{align}

\section{The commutator $[\hat{H}_I(t_1),\hat{H}_I(t_2)]$ in the Magnus expansion}
\label{appendixB}
Here, we show the calculation of the commutator $[\hat{H}_I(t_1),\hat{H}_I(t_2)]$ in the Magnus expansion.
The commutator reads
\begin{widetext}
\begin{equation}
\begin{aligned}
[\hat{H}_I(t_1),\hat{H}_I(t_2)]=&J(t_1)J(t_2)\sum_j \big[ e^{iU(\hat{O}_j-\hat{O}_{j+1}-1)t_1}e^{iU(\hat{O}_{j}-\hat{O}_{j+1}-2)t_2}(e^{-
iUt_2}-1)\hat{A}^{\dag}_{j}\hat{A}^{\dag}_j\hat{A}_{j+1}\hat{A}_{j+1}\\
&+ e^{iU(\hat{O}_j-\hat{O}_{j+1}-1)t_1}e^{iU(\hat{O}_{j+1}-\hat{O}_{j+2}-1)t_2}(e^{
iUt_2}-1)\hat{A}^{\dag}_{j}\hat{A}_{j+1}\hat{A}^{\dag}_{j+1}\hat{A}_{j+2} 
+ e^{iU(\hat{O}_j-\hat{O}_{j+1}-1)t_1}e^{iU(\hat{O}_{j+1}-\hat{O}_{j+2}-1)t_2}\hat{A}_{j+2}\hat{A}^{\dag}_{j}\\
&+ e^{iU(\hat{O}_j-\hat{O}_{j+1}-1)t_1}e^{iU(\hat{O}_j-\hat{O}_{j+1}-1)t_2}(e^{-iUt_2}-1)\hat{A}^{\dag}_{j}\hat{A}^{\dag}_{j}\hat{A}_{j+1}\hat{A}_{j+1}\\
&+ e^{iU(\hat{O}_j-\hat{O}_{j+1}-1)t_1}e^{iU(\hat{O}_{j-1}-\hat{O}_j-1)t_2}(e^{iUt_2}-1)\hat{A}^{\dag}_{j+1}\hat{A}_{j-1}\hat{A}^{\dag}_{j}\hat{A}_{j}
- e^{-iU(\hat{O}_{j-1}-\hat{O}_{j}-1)t_2}e^{iU(\hat{O}_j-\hat{O}_{j+1}-1)t_1}\hat{A}^{\dag}_{j-1}\hat{A}_{j+1}\\
&+ e^{iU(\hat{O}_{j}-\hat{O}_{j+1}-1)t_2}e^{iU(\hat{O}_j-\hat{O}_{j+1}-1)t_1}(1-e^{-iUt_1})\hat{A}^{\dag}_{j}\hat{A}^{\dag}_{j}\hat{A}_{j+1}\hat{A}_{j+1}
\\ &+e^{iU(\hat{O}_{j+1}-\hat{O}_{j+2}-1)t_2}e^{iU(\hat{O}_j-\hat{O}_{j+1}-1)t_1}(1-e^{iUt_1})\hat{A}^{\dag}_{j+1}\hat{A}_{j+1}\hat{A}_{j+2}\hat{A}^{\dag}_{j}\\
&+ e^{iU(\hat{O}_{j}-\hat{O}_{j+1}-1)t_2}e^{iU(\hat{O}_j-\hat{O}_{j+1}-2)t_1}(1-e^{iUt_1})\hat{A}^{\dag}_{j}\hat{A}^{\dag}_{j}\hat{A}_{j+1}\hat{A}_{j+1}
+ e^{iU(\hat{O}_{j-1}-\hat{O}_{j}-1)t_2}e^{iU(\hat{O}_j-\hat{O}_{j+1}-1)t_1}(1-e^{iUt_1})\hat{A}^{\dag}_{j-1}\hat{A}_{j}\hat{A}^{\dag}_{j}\hat{A}_{j+1}\\
&+ e^{iU(\hat{O}_{j+1}-\hat{O}_{j}-1)t_1}e^{iU(\hat{O}_{j+1}-\hat{O}_{j}-2)t_2}(e^{-iUt_2}-1)\hat{A}_{j}\hat{A}_{j}\hat{A}^{\dag}_{j+1}\hat{A}^{\dag}_{j+1}\\
&+ e^{iU(\hat{O}_{j+1}-\hat{O}_{j}-1)t_1}e^{iU(\hat{O}_{j+2}-\hat{O}_{j+1}-1)t_2}(e^{iUt_2}-1)\hat{A}_{j}\hat{A}^{\dag}_{j+1}\hat{A}_{j+1}\hat{A}^{\dag}_{j+2}\\
&- e^{iU(\hat{O}_{j+1}-\hat{O}_{j}-1)t_1}e^{iU(\hat{O}_{j+2}-\hat{O}_{j+1}-1)t_2}\hat{A}_{j}\hat{A}^{\dag}_{j+2}
+ e^{iU(\hat{O}_{j+1}-\hat{O}_{j}-1)t_1}e^{iU(\hat{O}_{j+1}-\hat{O}_{j}-1)t_2}(e^{-iUt_2}-1)\hat{A}_{j}\hat{A}_{j}\hat{A}^{\dag}_{j+1}\hat{A}^{\dag}_{j+1}\\
&+ e^{iU(\hat{O}_{j+1}-\hat{O}_{j}-1)t_1}e^{iU(\hat{O}_{j}-\hat{O}_{j-1}-1)t_2}(e^{iUt_2}-1)\hat{A}_{j-1}\hat{A}_{j}\hat{A}^{\dag}_{j}\hat{A}^{\dag}_{j+1} + e^{iU(\hat{O}_{j}-\hat{O}_{j-1}-1)t_2}e^{iU(\hat{O}_{j+1}-\hat{O}_j-1)t_1}\hat{A}_{j-1}\hat{A}^{\dag}_{j+1}\\
&+ e^{iU(\hat{O}_{j+1}-\hat{O}_{j}-1)t_2}e^{iU(\hat{O}_{j+1}-\hat{O}_j-1)t_1}(1-e^{-iUt_1})\hat{A}_{j}\hat{A}_{j}\hat{A}^{\dag}_{j+1}\hat{A}^{\dag}_{j+1}
+ e^{iU(\hat{O}_{j+2}-\hat{O}_{j+1}-1)t_2}e^{iU(\hat{O}_{j+1}-\hat{O}_j-1)t_1}(1-e^{iUt_1})\hat{A}^{\dag}_{j+2}\hat{A}_{j}\hat{A}_{j+1}\hat{A}^{\dag}_{j+1}\\
&+ e^{iU(\hat{O}_{j+1}-\hat{O}_{j}-1)t_2}e^{iU(\hat{O}_{j+1}-\hat{O}_j-2)t_1}(1-e^{-iUt_1})\hat{A}^{\dag}_{j+1}\hat{A}^{\dag}_{j+1}\hat{A}_j\hat{A}_{j}
+ e^{iU(\hat{O}_{j}-\hat{O}_{j-1}-1)t_2}e^{iU(\hat{O}_{j+1}-\hat{O}_j-1)t_1}(1-e^{iUt_1})\hat{A}_{j-1}\hat{A}^{\dag}_{j}\hat{A}_{j}\hat{A}_{j+1}\\
&+ e^{iU(\hat{O}_{j}-\hat{O}_{j+1}-1)t_1}e^{iU(\hat{O}_{j+1}-\hat{O}_j)t_2}(e^{iUt_2}-1)\hat{A}^{\dag}_{j}\hat{A}_{j}\hat{A}_{j+1}\hat{A}^{\dag}_{j+1}
+ e^{iU(\hat{O}_{j}-\hat{O}_{j+1}-1)t_1}e^{iU(\hat{O}_{j+2}-\hat{O}_{j+1}-1)t_2}(e^{-iUt_2}-1)\hat{A}^{\dag}_{j}\hat{A}_{j+1}\hat{A}_{j+1}\hat{A}^{\dag}_{j+2}\\
& e^{iU(\hat{O}_{j}-\hat{O}_{j+1}-1)t_1}e^{iU(\hat{O}_{j+1}-\hat{O}_j)t_2}\hat{A}^{\dag}_{j}\hat{A}_{j}
+ e^{iU(\hat{O}_{j}-\hat{O}_{j+1}-1)t_1}e^{iU(\hat{O}_{j+1}-\hat{O}_{j}-1)t_2}(e^{iUt_2}-1)\hat{A}^{\dag}_{j}\hat{A}_{j}\hat{A}^{\dag}_{j+1}\hat{A}_{j+1}\\
&+ e^{iU(\hat{O}_{j}-\hat{O}_{j+1}-1)t_1}e^{iU(\hat{O}_{j}-\hat{O}_{j-1}-1)t_2}(e^{-iUt_2}-1)\hat{A}_{j-1}\hat{A}^{\dag}_{j}\hat{A}^{\dag}_{j}\hat{A}_{j+1}
- e^{iU(\hat{O}_{j+1}-\hat{O}_{j}-1)t_2}e^{iU(\hat{O}_{j}-\hat{O}_{j+1}-1)t_1}\hat{A}^{\dag}_{j+1}\hat{A}_{j+1}\\
&+ e^{iU(\hat{O}_{j+1}-\hat{O}_{j}-1)t_2}e^{iU(\hat{O}_{j}-\hat{O}_{j+1}+2)t_1}(1-e^{-iUt_1})\hat{A}^{\dag}_{j+1}\hat{A}_{j+1}\hat{A}_{j}\hat{A}^{\dag}_{j}
+ e^{iU(\hat{O}_{j+2}-\hat{O}_{j+1}-1)t_2}e^{iU(\hat{O}_{j}-\hat{O}_{j+1}-1)t_1}(1-e^{-iUt_1})\hat{A}^{\dag}_{j}\hat{A}_{j+1}\hat{A}_{j+1}\hat{A}^{\dag}_{j+2}\\
&+ e^{iU(\hat{O}_{j+1}-\hat{O}_{j}-1)t_1}e^{iU(\hat{O}_{j}-\hat{O}_{j+1})t_2}(e^{iUt_2}-1)\hat{A}_{j}\hat{A}^{\dag}_{j}\hat{A}^{\dag}_{j+1}\hat{A}_{j+1}
+ e^{iU(\hat{O}_{j+1}-\hat{O}_{j}-1)t_2}e^{iU(\hat{O}_{j+1}-\hat{O}_{j+2}-1)t_2}(e^{-iUt_2}-1)\hat{A}_{j}\hat{A}^{\dag}_{j+1}\hat{A}^{\dag}_{j+1}\hat{A}_{j+2}\\
&- e^{iU(\hat{O}_{j+1}-\hat{O}_{j}-1)t_1}e^{iU(\hat{O}_{j}-O_{j+1})t_2}\hat{A}_{j}\hat{A}^{\dag}_{j}
+ e^{iU(\hat{O}_{j+1}-\hat{O}_{j}-1)t_1}e^{iU(\hat{O}_{j}-\hat{O}_{j+1}-1)t_2}(e^{iUt_2}-1)\hat{A}_{j}\hat{A}^{\dag}_{j}\hat{A}_{j+1}\hat{A}^{\dag}_{j+1}\\
&+ e^{iU(\hat{O}_{j}-\hat{O}_{j+1}-1)t_2}e^{iU(\hat{O}_{j+1}-\hat{O}_{j}-1)t_1}\hat{A}_{j+1}\hat{A}^{\dag}_{j+1}
+ e^{iU(\hat{O}_{j}-\hat{O}_{j+1}-1)t_2}e^{iU(\hat{O}_{j+1}-\hat{O}_{j})t_1}(e^{iUt_1}-1)\hat{A}^{\dag}_j\hat{A}_{j}\hat{A}_{j+1}\hat{A}^{\dag}_{j+1}\\
&+ e^{iU(\hat{O}_{j+1}-\hat{O}_{j+2}-1)t_2}e^{iU(\hat{O}_{j+1}-\hat{O}_{j}-2)t_1}(e^{iUt_1}-1)\hat{A}_{j+2}\hat{A}_{j}\hat{A}^{\dag}_{j+1}\hat{A}^{\dag}_{j+1}
+ e^{iU(\hat{O}_{j-1}-\hat{O}_{j}-1)t_2}e^{iU(\hat{O}_{j+1}-\hat{O}_{j}-2)t_1}(e^{iUt_1}-1)\hat{A}^{\dag}_{j-1}\hat{A}_{j}\hat{A}_{j}\hat{A}^{\dag}_{j+1} \big].
\label{commutator}
\end{aligned}
\end{equation}
\end{widetext}

\section{Analytical expressions for the trimer}
\label{appendixC}
The trimer dynamics allows us to compute analytical expressions for effective Hamiltonians and populations of quantum states. At the integer resonance $\Omega=U$, the wave function at time $t$ can be analytically computed by diagonalizing the effective Hamiltonian  $\hat{H}_F^{(0)}=-\hbar J_0(\ket{\psi_0}\bra{\psi_1}+\ket{\psi_0}\bra{\psi_2}+{\rm H.c})$. The quantum state can be written as 
\begin{equation}
\begin{aligned}
|\psi(t)\rangle=&\cos{(\sqrt{2}J_0t)}\ket{\psi_0} +\frac{i\sin{(\sqrt{2}J_0t)}}{\sqrt{2}}(\ket{\psi_1}+\ket{\psi_2}),
\label{WaveFunction}
\end{aligned}
\end{equation}
and the corresponding probabilities read
\begin{equation}
\begin{aligned}
P_0^{A}(t)=&\cos^2{(\sqrt{2}J_0t)}\\
P_1^{A}(t)=&\sin^2{(\sqrt{2}J_0t)}/2=P_2^{A}(t).
\label{Prob1}
\end{aligned}
\end{equation}
The main text plots these populations in Fig.~\ref{Fig4}(a).

At the fractional resonance, $\Omega=U/2$, second-order processes dominate the quantum dynamics via the effective $2\times 2$ Hamiltonian $\hat{H}_F^{(1)}=\frac{16\hbar J^2_0}{3U}\ket{\psi_0}\bra{\psi_0}+\frac{4\hbar J^2_0}{5U}\ket{\psi_3}\bra{\psi_3}+\frac{3\hbar J^2_0}{U}(\ket{\psi_0}\bra{\psi_3}+\ket{\psi_3}\bra{\psi_0})$. The quantum state at time $t$ reads
\begin{equation}
\begin{aligned}
|\psi(t)\rangle=&c_0(t)|\psi_0\rangle + c_1(t)|\psi_3\rangle,
\label{WaveFunction2}
\end{aligned}
\end{equation}
where the probability amplitudes are
\begin{equation}
\begin{aligned}
c_0(t)=&\frac{1}{4\lambda}[\left(2\lambda+a-c\right) e^{-\frac{1}{2} i t \left(2\lambda+a+c\right)}+\left(2\lambda-a+c\right) e^{\frac{1}{2} i t \left(2\lambda-a-c\right)}]\\
c_1(t)=&-\frac{ i b e^{-\frac{1}{2} i t (a+c)} \sin{\lambda t}}{\lambda},
\label{Amplitude}
\end{aligned}
\end{equation}
with $\lambda=\sqrt{(a-c)^2+4 b^2}/2$, $a=16J^2_0/3U$, $b=3J^2_0/U$ and $c=4J^2_0/5U$. The analytical populations simply read
\begin{equation}
\begin{aligned}
P_0^{A}(t)=&\frac{2 b^2 \cos \left( 2\lambda t \right)+(a-c)^2+2 b^2}{4 \lambda^2}\\
P_3^{A}(t)=&\frac{ b^2 \left(1-\cos \left(2\lambda  t \right)\right)}{2 \lambda^2}.
\label{Prob2}
\end{aligned}
\end{equation}
The main text plots these populations in Fig.~\ref{Fig4}(b).

\section{Performance of the time-evolving block decimation algorithm}
\label{appendixD}
\begin{figure}[t]
\centering
\includegraphics[scale=0.27]{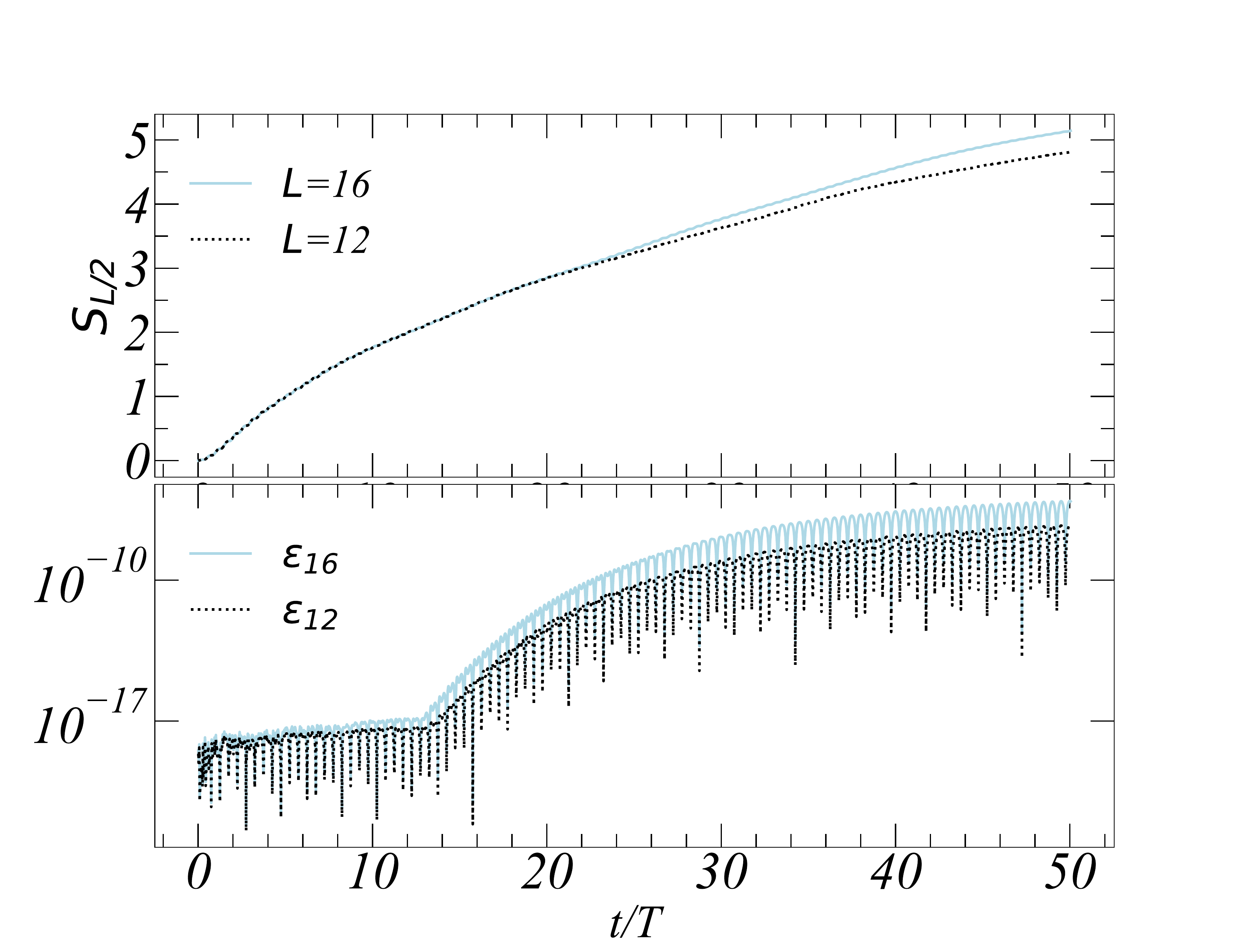}
\caption{Integer resonance case $\Omega=U$. The upper panel shows the half-chain von Neumann entropy for $L=12$ and $L=16$ lattice sites. The lower panel shows the truncation error $\epsilon$ in the logarithmic scale computed from the TEBD algorithm. In our numerical calculations we consider parameters $\omega=1$, $J_0= 0.01\omega$, and $U=40J_0$. We have truncated up to a maximum occupation number per site $n_{\rm max}=2$. We stress that increasing the maximum occupation number to $n_{\rm max}=3$ provides the same results, see Fig.~\ref{Fig16}.}
\label{Fig14}
\end{figure}

\begin{figure}[t]
\centering
\includegraphics[scale=0.27]{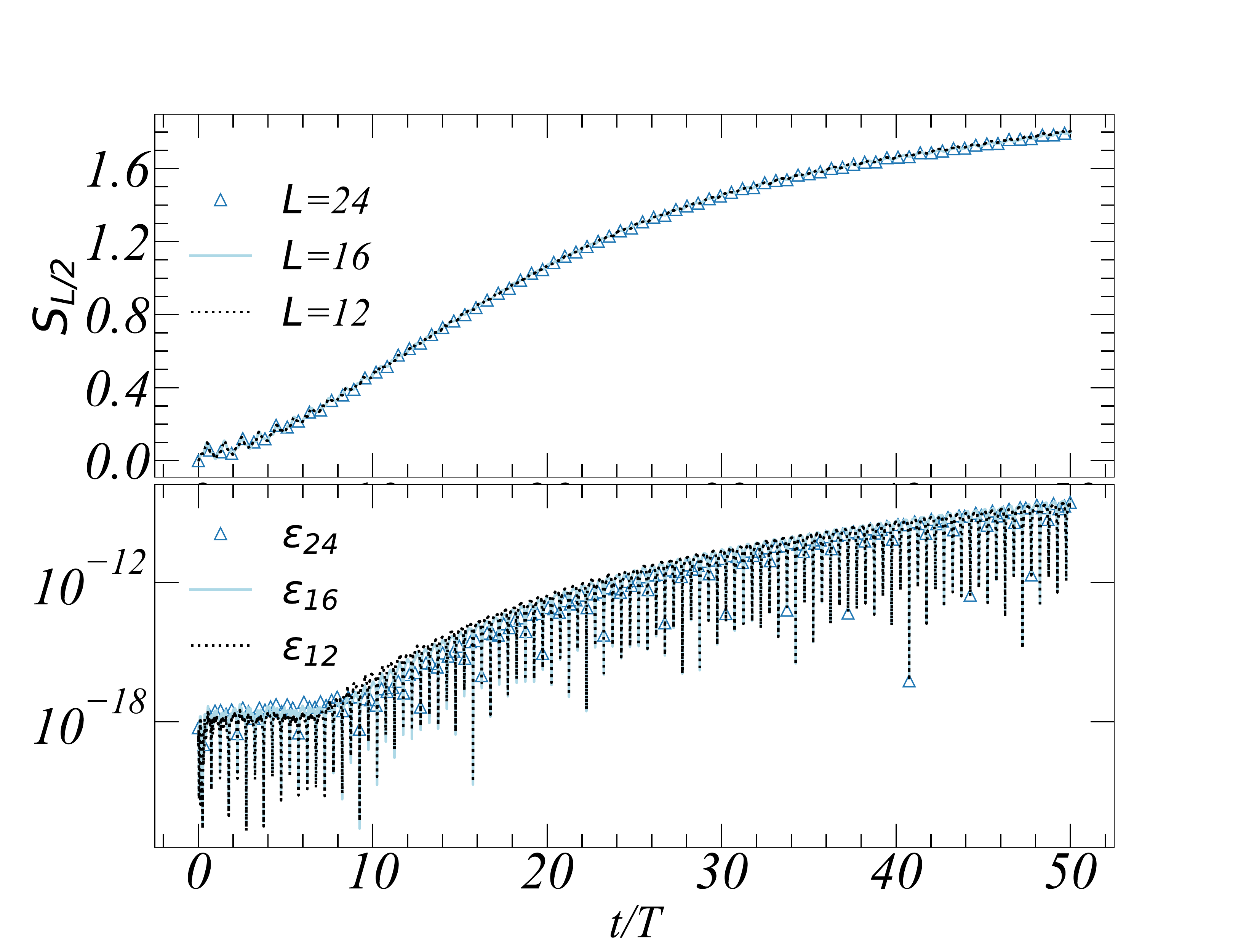}
\caption{Fractional resonance case $\Omega=U/2$. The upper panel shows the half-chain von Neumann entropy for $L=12$, $L=16$, and $L=24$ lattice sites. The lower panel shows the truncation error $\epsilon$ computed from the TEBD algorithm. In our numerical calculations we consider parameters $\omega=1$, $J_0= 0.01\omega$, and $U=40J_0$. We have truncated up to a maximum occupation number per site $n_{\rm max}=2$. We stress that increasing the maximum occupation number to $n_{\rm max}=3$ provides the same results, see Fig.~\ref{Fig16}.}
\label{Fig15}
\end{figure}

Here, we analyze the performance of the TEBD algorithm as we increase the system size. In particular, we consider the convergence of half-chain von Neumann entropy and the associated truncation error. In Fig.~\ref{Fig14}, we plot the half-chain von Neumann entropy (upper panel) and the truncation error (lower panel) in the integer resonance case, $\Omega=U$, for $L=12$ and $L=16$ lattice sites. In these simulations, we have checked the convergence of the half-chain von Neumann entropy for a large bond dimension of $\chi=520$. Notice that the truncation error remains very low within the simulating time, owing to the strongly interacting regime of the Bose-Hubbard model with parameters $U/J_0=40$. As we increase the lattice size, we see higher creation of bipartite entanglement over time due to the first-order processes that dominate the dynamics.  

In the fractional resonance, $\Omega=U/2$, we plot the half-chain von Neumann entropy (upper panel) and the truncation error (lower panel) for $L=12$, $L=16$, and $L=24$ lattice sites, see Fig.~\ref{Fig15}. These simulations confirmed the convergence of half-chain von Neumann entropy for moderate bond dimensions of $\chi=120$, $\chi=140$, and $\chi=180$, respectively. Notice that the truncation error remains low during the simulation due to the Bose-Hubbard model's strongly interacting regime with $U/J_0=40$ and the strong localization of the many-body quantum state. 
\begin{figure}[t]
\centering
\includegraphics[scale=0.17]{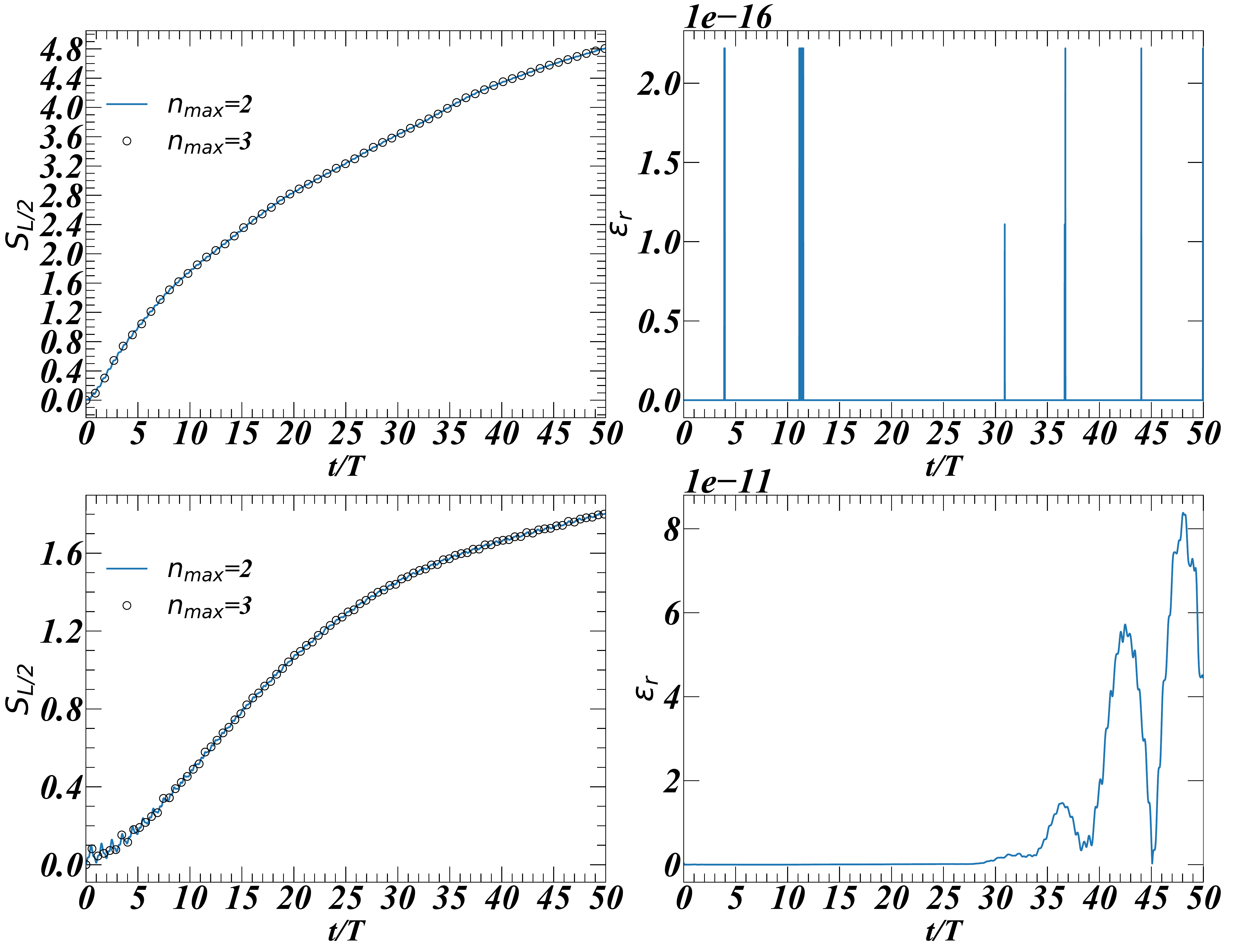}
\caption{Upper panel, we consider the integer resonance to compute the half-chain von Neumann entropy for a lattice of $L=12$ sites using a truncated local Hilbert space of $n_{\rm{max}}=2$ and $n_{\rm{max}}=3$, and their absolute relative error. Lower panel, we consider the fractional resonance to compute the half-chain von Neumann entropy for a lattice of $L=12$ sites using a truncated local Hilbert space of $n_{\rm{max}}=2$ and $n_{\rm{max}}=3$, and their absolute relative error. It is clear that truncating the local Hilbert space to $n_{\rm{max}}=2$ is enough to obtain trustful results.}
\label{Fig16}
\end{figure}

As we increase the lattice size, our results in Fig.~\ref{Fig15} provide clear evidence of the fractional resonance robustness. The robust localization of the many-body quantum state and the slowing down of the many-body quantum dynamics are still present. We observe the same increase in the half-chain von Neumann entropy over time. 

In the main text, we also state that numerical simulations using $n_{\rm{max}}=2$ and $n_{\rm{max}}=3$ provide the same results as we increase the lattice size. Figure~\ref{Fig16} offers clear evidence of our statement. Here, we show the half-chain von Neumann entropy as a function of time for a lattice size $L=12$ considering the integer resonance (upper panel) and the fractional resonance (lower panel). The right column shows their absolute relative errors.    

\section{Tensor product of algebras and Ladder systems}
\label{appendixE}
As a next category of models that exhibit integer and fractional resonances, we consider spin ladder systems. For simplicity, let's consider the tensor product $\mathcal{G}\otimes \mathcal{G}$ where $\mathcal{G}=\mathcal{H}\bigoplus_\alpha \mathcal{G}_\alpha$, where $\mathcal{H}$ is the Cartan subalgebra that we discussed in Sec.~\ref{sec:II}.

Now, let us consider two generators of the Cartan algebra that we call $\hat{O}_{i,a}$ and $\hat{O}_{i,b}$. Consequently, we consider local ladder operators $\hat{A}_{j,a}$ and $\hat{A}^{\dagger}_{j,b}$ such that they satisfy the algebraic relations $[\hat{O}_{i,a},\hat{A}^{\dagger}_{j,b}]=\alpha\delta_{i,j}\delta_{a,b}\hat{A}^{\dagger}_{j,b}$, and $[\hat{O}_{i,a},\hat{A}_{j,b}]=-\alpha\delta_{i,j}\delta_{a,b}\hat{A}_{j,b}$, where $\alpha$ is a real constant. Here the indices $a$ and $b$ act as "flavors" and allows us to distinguish the different algebras. By using this notation, we can write the generic ladder Hamiltonian
\begin{align}
          \label{eq:HamAlgebraLadder}
                  \hat{H}&=\hbar\sum^L_{j=1,\theta={a,b}}\omega \hat{O}_{j,\theta}+\hbar U\sum^L_{j=1}\hat{O}_{j,a}\hat{O}_{j,b}
 \nonumber\\&
 -\hbar J_0\cos{\Omega t}\sum^{L-1}_{j=1,\theta={a,b}}(\hat{A}^{\dagger}_{j,\theta}\hat{A}_{j+1,\theta}+\hat{A}^{\dagger}_{j+1,\theta}\hat{A}_{j,\theta})
 \ .
\end{align}
This Hamiltonian describes the dynamics of a two-leg ladder. Each leg is labelled by an index $\theta=a,b$. The coupling between the legs is $\hat{O}_{j,a}\hat{O}_{j,b}$.
Now we follow similar steps as before, but now we go to the rotating frame with the Hamiltonian $\hat{H}_0=U\sum^N_{j=1}\hat{O}_{j,a}\hat{O}_{j,b}
$. In the rotating frame, we will find expressions like
\begin{equation}
          \label{eq:OperatorEvLadder}
                  \hat{A}^{\dagger}_{j,b}(t)=e^{\frac{i}{\hbar}\hat{H}_0t}\hat{A}^{\dagger}_{j,b}e^{-\frac{i}{\hbar}\hat{H}_0t}
                  \ .
\end{equation}
Again, by taking the derivative of this expression, we end up with terms like
\begin{equation}
          \label{eq:OperatorEvLadder}
                  \frac{d \hat{A}^{\dagger}_{j,b}(t)}{dt}=i U[\hat{O}_{j,a}\hat{O}_{j,b},\hat{A}^{\dagger}_{j,b}(t)]=i U \alpha\hat{O}_{j,a}\hat{A}^{\dagger}_{j,b}(t)
                  \ .
\end{equation}
From this we can get the time evolution $\hat{A}^{\dagger}_j(t)=\exp(iU\alpha t\hat{O}_{j,a})\hat{A}^{\dagger}_j$. Similarly, we can proof that $\hat{A}_j(t)=\exp(-iU\alpha t\hat{O}_{j,a})\hat{A}_j$. By using this, we obtain the explicit form for the Hamiltonian in the rotating frame
\begin{align}
          \label{eq:OperatorEvLadder}
                  \hat{H}_{I}(t)&=e^{\frac{i}{\hbar}\hat{H}_0t} \hat{H}e^{-\frac{i}{\hbar}\hat{H}_0t} \nonumber\\
                  &=-\hbar J_0\cos{\Omega t}\sum^{L-1}_{j=1}(e^{iU \alpha t(\hat{O}_{j+1,b}-\hat{O}_{j,b})}\hat{A}^{\dagger}_{j,a}\hat{A}_{j+1,a}+\rm{H.c}                 
                   \nonumber\\&
                   -\hbar J_0\cos{\Omega t}\sum^{L-1}_{j=1}(e^{iU \alpha t(\hat{O}_{j+1,a}-\hat{O}_{j,a})}\hat{A}^{\dagger}_{j,b}\hat{A}_{j+1,b}+\rm{H.c}
                  \ .
\end{align}
Here let us define $\pm m_{\Omega}=\pm (m_{j+1,\theta}-m_{j,\theta})$. Fractional resonances will occur whenever the condition $m_{\Omega}=2\Omega/U\alpha$ is satisfied. This depends intimately on the relation between the consecutive quantum numbers $m_{j+1}$ and $m_{j}$ and the nature of the local Hilbert space at the $j$th site. Note that here the local Hermitian operator $\hat{O}_{j,\theta}$ satisfies the eigenvalue equation $\hat{O}_{j,\theta}\ket{m_{j,\theta}}=\alpha m_{j,\theta}\ket{m_{j,\theta}}$, where $m_{j,\theta}$ is an integer number and $\theta=a,b$ denotes the two "flavors."
\subsection{Spin ladders and relation to Fermi-Hubbard model}
Now it is to consider an example of the general theory. With this aim, let us consider diagonal spin operators $\hat{\sigma}^z_{i,\theta}$ with "two flavors" $\theta=a,b$ denoting a given spin chain labeled by a or b. Using this construction, we can define the local operator $ \hat{O}_{i,\theta}=\hat{\sigma}^z_{i,\theta}$. Based on the general algebraic construction, we can define the ladder Hamiltonian

\begin{equation}
          \label{eq:HamAlgebra1}
                  \hat{H}=\hbar\sum^L_{j, \theta}\omega \hat{\sigma}^z_{i,\theta}+\frac{\hbar U}{2}\sum^L_{j}\hat{\sigma}^z_{i,a}\hat{\sigma}^z_{i,b}-\hbar J_0\cos{\Omega t}\sum_{j, \theta}(\hat{\sigma}^+_{i,\theta}\hat{\sigma}^{-}_{i+1,\theta}+\rm{H.c})
\end{equation}

After a Jordan Wigner transformation, this model maps to the Fermi-Hubbard Hamiltonian.

\begin{equation}
          \label{eq:HamAlgebra1}
                  \hat{H}=\hbar\sum^L_{j, \theta}\tilde{\omega}\hat{f}^{\dagger}_{i,\theta}\hat{f}_{i,\theta}+\frac{\hbar U}{2}\sum^L_{j}\hat{f}^{\dagger}_{i,a}\hat{f}_{i,a}\hat{f}^{\dagger}_{i,b}\hat{f}_{i,b}-\hbar J_0\cos{\Omega t}\sum_{j, \theta}(\hat{f}^{\dagger}_{i,\theta}\hat{f}_{i+1,\theta}+\rm{H.c}),
\end{equation}
where $\tilde{\omega}=\omega-U$.

\bibliography{Mybib}

\end{document}